\documentclass[prb,aps,twocolumn,superscriptaddress,showpacs]{revtex4-1}
\usepackage[colorlinks=true,linkcolor=blue, citecolor=blue, urlcolor=blue, unicode=true]{hyperref}
\usepackage{graphicx}
\usepackage{amsmath,amssymb} 
\usepackage{physics}
\usepackage{chemformula}
\usepackage{siunitx}
\usepackage{tabularx}
\usepackage[capitalise]{cleveref} 

\newcommand\ut[1]{_\mathrm{#1}} 

\newcommand\et{\ensuremath{\epsilon\ut{T}}}
\newcommand\eh{\ensuremath{\epsilon\ut{H}}}
\newcommand\etc{\ensuremath{\eta\ut{C}}}
\newcommand\ett{\ensuremath{\eta\ut{T}}}
\newcommand\mnsb{\ch{MnSb2O6}}
\newcommand\bnfs{\ch{Ba3NbFe3Si2O14}}
\newcommand\tn{\ensuremath{T\ut{N}}}
\newcommand\Q{\ensuremath{\vb*{Q}}}
\newcommand\Mperp{\vb*{M}_{\perp}}
\newcommand\Mp[1]{M_{\perp #1}}

\newcommand\ii{\mathrm{i}}
\newcommand\ee{\mathrm{e}}

\begin{document}

\title{Neutron diffraction in \mnsb:  Magnetic and structural domains in a helicoidal polar magnet with coupled chiralities}

\author{E. Chan}
\affiliation{Institut Laue-Langevin, 71 avenue des Martyrs, CS 20156, 38042 Grenoble Cedex 9, France}
\affiliation{School of Physics and Astronomy, University of Edinburgh, Edinburgh EH9 3JZ, United Kingdom}

\author{J. P\'{a}sztorov\'{a}}
\affiliation{School of Physics and Astronomy, University of Edinburgh, Edinburgh EH9 3JZ, United Kingdom}

\author{R. D. Johnson}
\affiliation{Department of Physics and Astronomy, University College London, Gower Street, London WC1E 6BT}

\author{M. Songvilay}
\affiliation{School of Physics and Astronomy, University of Edinburgh, Edinburgh EH9 3JZ, United Kingdom}

\author{R. A. Downie}
\affiliation{Institute of Chemical Sciences and Centre for Advanced Energy Storage and Recovery, School of Engineering and Physical Sciences, Heriot-Watt University, Edinburgh EH14 4AS, United Kingdom}

\author{J-W. G. Bos}
\affiliation{Institute of Chemical Sciences and Centre for Advanced Energy Storage and Recovery, School of Engineering and Physical Sciences, Heriot-Watt University, Edinburgh EH14 4AS, United Kingdom}

\author{O.~Fabelo}
\affiliation{Institut Laue-Langevin, 71 avenue des Martyrs, CS 20156, 38042 Grenoble Cedex 9, France}

\author{C. Ritter}
\affiliation{Institut Laue-Langevin, 71 avenue des Martyrs, CS 20156, 38042 Grenoble Cedex 9, France}

\author{K. Beauvois}
\affiliation{Institut Laue-Langevin, 71 avenue des Martyrs, CS 20156, 38042 Grenoble Cedex 9, France}

\author{Ch. Niedermayer}
\affiliation{Laboratory for Neutron Scattering, Paul Scherrer Institut, CH-5232 Villigen, Switzerland}

\author{S.-W. Cheong}
\affiliation{Rutgers Center for Emergent Materials and Department of Physics and Astronomy, Rutgers University, 136 Frelinghuysen Road, Piscataway, New Jersey 08854, USA}

\author{N. Qureshi}
\affiliation{Institut Laue-Langevin, 71 avenue des Martyrs, CS 20156, 38042 Grenoble Cedex 9, France}

\author{C. Stock}
\affiliation{School of Physics and Astronomy, University of Edinburgh, Edinburgh EH9 3JZ, United Kingdom}

\date{\today}

\begin{abstract}
	
MnSb$_{2}$O$_{6}$ is based on the structural chiral $P$321 space group \#150 where the magnetic Mn$^{2+}$ moments ($S=5/2$, $L\approx 0$) order antiferromagnetically at $T\ut{N}=12$ K. Unlike the related iron based langasite (\bnfs) where the low temperature magnetism is based on a proper helix characterized by a time-even \emph{pseudo}scalar `magnetic' chirality, the Mn$^{2+}$ ions in MnSb$_{2}$O$_{6}$ order with a cycloidal structure at low temperatures, described instead by a time-even vector `magnetic' polarity.  A tilted cycloidal structure has been found [M. Kinoshita \textit{et al.} Phys. Rev. Lett. {\bf{117}}, 047201 (2016)] to facilitate ferroelectric switching under an applied magnetic field.  In this work, we apply polarized and unpolarized neutron diffraction analyzing the magnetic and nuclear structures in MnSb$_{2}$O$_{6}$ with the aim of understanding this magnetoelectric coupling. We find no evidence for a helicoidal magnetic structure with one of the spin envelope axes tilted away from the cycloidal $c$-axis. 
However, on application of a magnetic field $\parallel$ $\vb*{c}$ the spin rotation plane can be tilted, giving rise to a cycloid---helix admixture that evolves towards a distorted helix (zero cycloidal component) for fields great than $\approx$ 2 T.  We propose a mechanism for the previously reported ferroelectric switching based on coupled structural and magnetic chiralities requiring only an imbalance of structural chiral domains.

\end{abstract}

\pacs{}

\maketitle

\section{Introduction}

Coupling magnetism and ferroelectricity would allow the possibility for controlling electric polarization with a magnetic field and magnetic moments with an electric field. However, ferroelectricity and magnetism originate from disparate microscopic mechanisms\cite{spaldin2005309}, and such multiferroic materials are rare. Despite these challenges, complex coupling schemes have been intensively studied and sought after for decades, motivated by the interesting physics and promising multifunctional applications.\cite{cheong20076,tokura201477,dong201564,fiebig20161a}  For example, non-centrosymmetric magnetic ordering can break inversion symmetry and induce an improper electric polarization via the inverse antisymmetric Dzyaloshinskii-Moriya (DM) interaction.\cite{kimura2003426,kenzelmann200595} This is the case in cycloidal magnets, often stabilized by the competition of exchange interactions, and where the sense of rotation of the spins can be linked to the sign of the electric polarization.\cite{katsura200595,mostovoy200696,kimura200737} 
Additional interest can be found in materials having a crystallographic chirality that may naturally stabilize a non-centrosymmetric magnetic structure. For example, iron based langasite (\bnfs\ [\onlinecite{marty2008101,loire2011106,stock201183,chaix2016,stock2019100}]) crystallizes in the chiral, trigonal space group $P$321, and the structural chirality is coupled to the chirality of its magnetic helix through symmetric Heisenberg exchanges. Recently, a magnetic field induced long-wavelength spin spiral modulation has been discovered in this compound giving rise to an electric polarization.\cite{ramakrishnan20194} 

\begin{figure}[h]
    \begin{center}
		\includegraphics[scale=0.3]{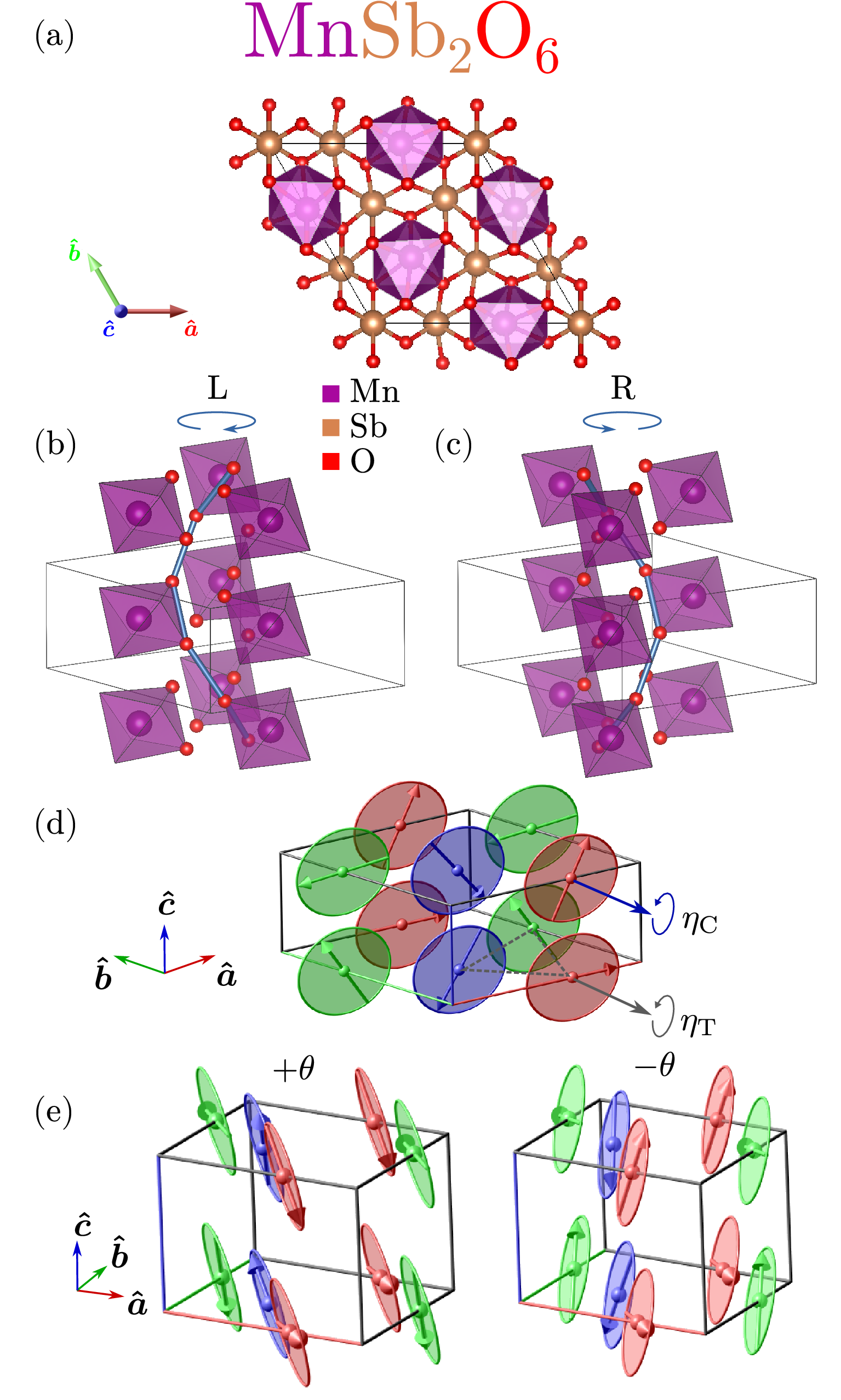}    
	\end{center}
    \caption{(a) Nuclear structure of chiral \mnsb. The structural chirality can be defined as the helical winding of the Mn-O-O-Mn super-super-exchange (blue lines) with respect to the $c$-axis: it is clockwise for left-handed structure (b) and anticlockwise for right-handed structure (c). Figures made on \textsc{Vesta}.\cite{momma201144} (d) Cycloidal magnetic structure. (e) Tilted cycloid model. Figures made on \textsc{Mag2Pol}.\cite{qureshi201952}}
    \label{fig:MSO_struc}
\end{figure}

\mnsb\ hosts magnetic manganese with a valence of Mn$^{2+}$ giving a high spin $S=5/2$ and orbitally quenched $L\approx 0$ moment.\cite{johnson2013111}  Analogous to iron based langasite, magnetic interactions between isolated \ch{MnO6} octahedra  (\cref{fig:MSO_struc}(a)) follow chiral super-super-exchange (SSE) pathways (Mn-O-O-Mn) along the $c$-axis (\cref{fig:MSO_struc}(b)-(c)). Magnetization measurements find long-range magnetic order below $\tn\approx\SI{12}{K}$ and some evidence for short-range correlations below $\SI{200}{K}$ has been provided.\cite{reimers1989,werner201694} Below $\tn$, Mn magnetic moments rotate within the $(ac)$-plane. Nearest neighbor moments arranged on triangular motifs in the $(ab)$-plane are dephased by 120° and follow a cycloidal modulation with propagation vector $\vb*{k} = (0,0,0.182)$, as shown in \cref{fig:MSO_struc}(d). The sense of rotation of the spins along the $c$-axis and within a basal triangle can be described by so-called vector chiralities, $\vb*{V}_\mathrm{C}$ and $\vb*{V}_\mathrm{T}$, respectively, which we later show to be related to well-defined, generic `magnetic' parameters $\eta_\mathrm{C}$ and $\eta_\mathrm{T}$ that couple directly to the crystal chirality, $\sigma$. By analogy with various other cycloidal magnets,\cite{katsura200595,mostovoy200696} \mnsb\ can hold an electric polarization. 

Magnetic domains can exist when the symmetry of the paramagnetic phase is lowered by the ordered magnetic structure. These domains are energetically equivalent, and related by the symmetry operators which are broken during the phase transition.\cite{aizu19702} In the case of \mnsb, threefold symmetry is broken by the cycloidal magnetic structure, hence at least three cycloidal domains are expected below \tn. Additional magnetic domains related to the signs of $\eta_\mathrm{C}$ and $\eta_\mathrm{T}$ will also form, as discussed later. These magnetic domains are polar, so could be manipulated by an external electric field.\cite{johnson2013111}  Later on, Kinoshita \textit{et al.} found the cycloids to be tilted away from the $c$-axis, with one of the main axes of the spin envelope parallel to $[1\bar{1}0]$,\cite{kinoshita2016117} as shown in \cref{fig:MSO_struc}(e). The tilt of this alternative ground state magnetic structure was reported to be necessary to explain a macroscopic electric polarization evidenced by measuring pyroelectric current, confirming the multiferroic character of \mnsb.

In this paper, we apply unpolarized and polarized neutron diffraction to show that there is no clear evidence of this tilted model for the magnetic ground state. We do find evidence for a mixture of chiral structural domains in our single-crystal. Through magnetic diffraction under an applied magnetic field, we show that it is possible to manipulate the magnetic structure with small magnetic fields. Finally we propose an alternative mechanism for the appearance of electric polarization, based on the DM interaction under an external magnetic field and coupled chiralities.  This mechanism does not require a tilted cycloid ground state for ferroelectric domain switching in an applied magnetic field.

This paper is based upon five sections including this introduction.  After describing the materials preparation and neutron instrumentation used for diffraction studies in \cref{sec:exp}, we define twinning afforded by the $P$321 symmetry and various structural and magnetic chiralities  in MnSb$_{2}$O$_{6}$ in \cref{sec:theory}. In \cref{sec:results}, we describe the experimental results and finish in \cref{sec:mf} with a phenomenological theory for ferroelectric switching previously observed.   

\section{Experimental details}\label{sec:exp}

In this section we describe the materials preparation and neutron scattering experiments used to study both powders and single crystals of MnSb$_{2}$O$_{6}$.

\subsection{Materials preparation}\label{ssec:preparation}

Materials preparation followed the procedure outlined in Ref.~\onlinecite{nakua1995154}. Powders of MnSb$_{2}$O$_{6}$ were prepared by mixing stoichiometric amounts of pure MnCO$_{3}$ and Sb$_{2}$O$_{3}$. After mixing through grinding, the powder was pressed into a pellet and heated up to 1000$^{\circ}$C with the process repeated with intermediate grinding. It was found that heating the pellet to higher temperatures introduced the impurity Mn$_{2}$Sb$_{2}$O$_{7}$. Single crystals of MnSb$_{2}$O$_{6}$ were prepared using the flux method. Starting ratios for single-crystal growth were (by weight) 73\% of flux V$_{2}$O$_{5}$, 20\% of polycrystalline MnSb$_{2}$O$_{6}$ and 7\% of B$_{2}$O$_{3}$. The powder was ground and pressed into a pellet and flame sealed in a quartz ampoule under vacuum (less than 1e$^{-4}$ Torr). B$_{2}$O$_{3}$ was used to lower the melting temperature of the V$_{2}$O$_{5}$ flux. Back filling the ampoules with $\approx$ 200 mTorr of Argon gas was found to noticeably improve crystal sizes.  Quartz ampoules were then heated to 1000$^{\circ}$C at a rate of 60$^{\circ}$C/hour and soaked at this temperature for 24 hours. The furnace was then cooled to 700$^{\circ}$C at a rate of 2$^{\circ}$C/hour and held for 24 hours, before it was switched off and allowed to cool to room temperature. Crystal sizes in the range from a few millimeters to nearly a centimeter were obtained through this procedure. 

\subsection{Neutron diffraction} 

The nuclear and magnetic structures of \mnsb\ were studied on the four-circle diffractometers D9~[\onlinecite{MSO2021D9}] and D10~[\onlinecite{MSO2021D10}] (ILL, Grenoble) using a single crystal sample of dimensions $\sim 3\times 2\times 0.2$ mm$^3$ (hexagonal shape). On D9, a monochromatic neutron beam of wavelength $\lambda=\SI{0.836}{\angstrom}$ was selected by the (220) reflection of a Cu monochromator in transmission geometry. On D10, a wavelength of $\lambda=\SI{2.36}{\AA}$ was selected from a vertically focusing pyrolytic graphite monochromator. The same single crystal was previously characterized using the CRYOgenic Polarization Analysis Device (CRYOPAD)\cite{tasset1999267-268a} on the spin-polarized hot neutron diffractometer D3~[\onlinecite{MSO2017D3}] (ILL, Grenoble) using a wavelength $\lambda=\SI{0.85}{\AA}$ selected by the (111) reflection of a \ch{Cu2MnAl} Heusler monochromator. The good quality of the single crystal was confirmed by neutron Laue diffraction.
Powder diffraction was performed on the high-intensity two-axis diffractometer D20~[\onlinecite{MSO2021D20}] (ILL, Grenoble) on $\sim \SI{17}{g}$ of powder, using a wavelength $\lambda=\SI{2.41}{\AA}$ selected by the (002) reflection of a pyrolitic graphite HOPG monochromator in reflection position. single crystal diffraction under an external magnetic field was performed on the cold triple-axis spectrometer  RITA-2 (now replaced by CAMEA, SINQ, Villigen), using a horizontal cryo-magnet MA7 with wavelength $\lambda=\SI{4.9}{\AA}$ monochromated with a vertically focused pyrolitic graphite PG002 monochromator.  The use of a horizontal field was necessary given the need to apply the magnetic field along the $c$-axis, parallel to the magnetic propagation vector which is kinematically constrained to be in the horizontal plane.

While conventional powder and single crystal neutron diffraction was used in this work, we relied as well heavily on the use of less standard techniques: Schwinger scattering and spherical neutron polarimetry, to gain extra information into the complex nuclear and magnetic structures of MnSb$_{2}$O$_{6}$.  We briefly outline the theory of these techniques here before discussing the structural properties specific to MnSb$_{2}$O$_{6}$.

\subsubsection{Schwinger scattering for Structural Handedness}

In the reference frame of a moving neutron, the electric field of a non-centrosymmetric crystal creates an effective magnetic field which couples to the neutron spin. This neutron spin-orbit interaction results in a polarization-dependant scattering known as Schwinger scattering\cite{schwinger194873} which can be used as a probe of the structural handedness of the crystal.\cite{felcher197531,qureshi2020102} 

In the local coordinates, where $\vu*{z}\parallel\vb*{k}\ut{i}\times\vb*{k}\ut{f}$ is perpendicular to the scattering plane, the asymmetric Schwinger structure factor is given by:\cite{qureshi2020102}

\begin{equation}
F\ut{SO}(\Q) = \ii\frac{\gamma r_0}{2}\frac{m\ut{e}}{m\ut{p}}F\ut{E}(\Q)\cot(\theta)\vu*{\sigma}\cdot\vu*{z}
\end{equation}

\noindent where $\gamma$ is the neutron gyromagnetic ratio, $r_0$ is the electron classical radius, $\theta$ is the scattering angle, $\vu*{\sigma}$ is the neutron spin operator, and $F\ut{E}(\Q)$ is the electrostatic unit cell structure factor:

\begin{equation}
F\ut{E}(\Q)=\sum_j [Z_j-f_j(\Q)]\ee^{-W_j(\Q)}\ee^{-\ii\Q\cdot\vb*{r}_j}
\end{equation}

\noindent with $Z_j$, $f_j(\Q)$ and $W_j(\Q)$, respectively the atomic number, the X-ray atomic form factor, and the Debye-Waller factor of the $j$-th atom of the unit cell. The small ratio between the electron and proton mass $m\ut{e}/m\ut{p}$ leads to a weak Schwinger scattering cross-section ${\propto \frac{\gamma r_0}{2}\frac{m\ut{e}}{m\ut{p}}=\SI{-1.46e-4}{}}$ in units of $\SI{e-12}{cm}$ (nuclear scattering length).\cite{felcher197531} For a nuclear reflection, the contribution from Schwinger scattering adds to the nuclear structure factor $F\ut{N}$ leading to an intensity:

\begin{equation}
I^{\pm}\propto \abs{F\ut{N}}^2+\abs{F\ut{SO}}^2\pm\mathcal{I}
\end{equation}

\noindent where $\mathcal{I}=2p\Re(F\ut{N}F\ut{SO}^*)$ is an interference term and $p$ is the polarization of the incident beam along $\pm\vu*{z}$. Measuring both intensities with the incident neutron beam polarized along $\pm\vu*{z}$ allows us to compute the flipping ratio:

\begin{equation}
R =\frac{\abs{F\ut{N}}^2+\abs{F\ut{SO}}^2+\cal{I}}{\abs{F\ut{N}}^2+\abs{F\ut{SO}}^2-\cal{I}}.
\end{equation}

\noindent The flipping ratio technique, very well known to the magnetization density community, affords the extraction of the weak Schwinger scattering and to distinguish structural twins in a single crystal discussed below. Indeed, each twin would lead to a different flipping ratio. For example the flipping ratio is inverted for an inversion twin. 

\subsubsection{Spherical Neutron Polarimetry for Magnetic Handedness}

Spherical neutron polarimetry (SNP) is a powerful technique used to determine complex magnetic structures and magnetic chiralities (an example illustrated in Ref. \onlinecite{giles-donovan2020102}), which plainly demonstrates the benefits of using polarized neutrons.\cite{BROWN2006215,simonet2012213} The idea is to measure in three orthogonal directions the final polarization of the neutrons, for incident neutrons polarized along each of the three directions. This allows to measure a $3\times3$ polarization matrix given by:

\begin{equation}
P_{if}=\frac{n_{if} - n_{i\bar{f}}}{n_{if} + n_{i\bar{f}}}
\end{equation}

\noindent where $i,f=x,y,z$ denotes the polarization direction of the incident and scattered neutrons in the local coordinates where $\vu*{x}$ is parallel to the scattering vector, $\vu*{z}$ is perpendicular to the scattering plane and $\vu*{y}$ completes this right-handed set, $n_{if}$ and $n_{i\bar{f}}$ are the number of scattered neutrons with spin parallel and antiparallel to $f$-direction. The theoretical cross-sections for polarized neutrons are given by the Blume-Maleev equations.\cite{blume1963130,maleev19634} For a purely elastic magnetic reflection and a perfectly polarized neutron beam the polarization matrix $P_{if}$ is given by:

\scriptsize
\begin{equation}
\begin{pmatrix} 
\dfrac{-\abs*{\Mperp}^2-{M\ut{ch}}}{\abs*{\Mperp}^2+ M\ut{ch}} & 0 & 0\\ 
\dfrac{-{M_{ch}}}{\abs*{\Mperp}^2} & \dfrac{(\abs*{\Mp{y}}^2-\abs{\Mp{z}}^2)}{\abs{\Mperp}^2} & \dfrac{2\Re{\Mp{y}\Mp{z}^*}}{\abs{\Mperp}^2}\\ 
\dfrac{-{M_{ch}}}{\abs*{\Mperp}^2} & \dfrac{2\Re{\Mp{z}\Mp{y}^*}}{\abs{\Mperp}^2}& \dfrac{(\abs*{\Mp{z}}^2-\abs*{\Mp{y}}^2)}{\abs{\Mperp}^2}
\end{pmatrix}
\end{equation}
\normalsize

\noindent While unpolarized single crystal diffraction is only sensitive to the amplitude squared of the magnetic interaction vector $\abs{\Mperp}^2$, its components are accessible through SNP. Also,  $M\ut{ch}=2\Im{\Mp{y}\Mp{z}^*}$ gives information on the magnetic chirality.

\section{Theory and definitions}\label{sec:theory}

Given the complexity of the magnetic and nuclear structure in MnSb$_{2}$O$_{6}$, we outline in this section the various definitions for the structural and magnetic chiralities and twins.  This is required for presenting powder and single crystal neutron diffraction results discussed below.

\subsection{Definition of Twins}

Twinning occurs when two or more single crystals of the same species are intergrown in different orientations, related by the so-called twin laws.\cite{hahn2013D,parsons200359}
When the twin operation belongs to the point group of the lattice but not to the point group of the crystal, the twinning is called twinning by merohedry. In this case, the crystal lattices of the two twins overlap in both direct and reciprocal space.\cite{koch2006C} As all Bravais lattices are centrosymmetric, the non-centrosymmetric basis of \mnsb\ (space group $P321$) is expected to form inversion twins. Furthermore, the absence of improper rotations in $P321$ (\emph{e.g.} mirror plane) implies the inversion twins will have opposite structural chiralities (known as enantiomorphs). It follows that the reciprocal lattice of one twin is the inverse of the other, \emph{i.e.} $(hkl) \rightarrow (\bar{h}\bar{k}\bar{l})$. In the case of the $P$321 space group additional merohedral twinning associated with twofold rotation around the $c$-axis, \emph{i.e.} $(hkl) \rightarrow (\bar{h}\bar{k}l)$, is also allowed.\cite{chandra199955} We note that these twins related by twofold rotation have the same chirality. Combining the twofold rotation with the inversion twin leads to a fourth twin $(hk\bar{l})$.  In order to distinguish the structural chirality of these four possible merohedral twins, we will subsequently use the labels L$(hkl)$, L$(\bar{h}\bar{k}l)$, R$(\bar{h}\bar{k}\bar{l})$ and R$(hk\bar{l})$, where L(R) refers to the left(right)-handedness of the crystal structure, defined by the helical winding of the Mn-O-O-Mn super-super-exchange pathways in \cref{fig:MSO_struc}(b)-(c).

\subsection{Definition of Structural and Magnetic Chiralities}\label{sec:chira}

In crystallography, chirality can be defined as the property of an object ``being non-superposable by pure rotation and translation on its image formed by inversion through a point".\cite{flack200386} On the other hand, the definition of magnetic chirality is not obvious because the time reversal operation (T) has to be considered in addition to parity operation (P). Barron proposed a more general definition: ``True chirality is possessed by systems that exist in two distinct enantiomeric states that are interconverted by space inversion but not by time reversal combined with any proper spatial rotation."\cite{barron1986108} In this meaning only helical magnetic structures are truly chiral.\cite{simonet2012213,johnson201444} However, spin ``chirality" is commonly used to refer to the sense of rotation of the spins with respect to a crystallographic reference often taken to be an oriented link between two atomic sites, say $\vb*{r}_{ij}$, and can thus describe the spin configuration of cycloidal structures and triangular networks.\cite{villain197710} 

The cross-product of two spins at sites $i$ and $j$ defines a vector chirality

\begin{equation}
\vb*{V}_{ij}=\vb*{S}_i\times \vb*{S}_j
\end{equation}

\noindent which is a T-even axial vector (\emph{i.e.} P-even), changing sign on exchange of indices $i\leftrightarrow j$. This chirality vector is well-defined by providing the oriented link between two spins.

\begin{figure}[h]
    \begin{center}
		\includegraphics[scale=1]{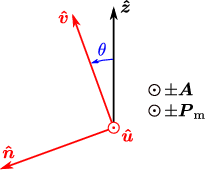}    
	\end{center}
    \caption{$\vu*{u}$ and $\vu*{v}$ are the main axis of the helicoidal spin structure envelope. Any cross product of spins lies along $\vu*{n}$. $\vb*{A}$ and $\vb*{P}\ut{m}$ lie along $\vu*{u}$.}
    \label{fig:basis}
\end{figure}

For clarification and to understand our diffraction data, we redefine the vectors introduced in Ref.~\onlinecite{johnson2013111} in the context of \mnsb. To do this, we consider an orthonormal basis $\mathsf{R}~=~(\vu*{x},\vu*{y},\vu*{z})$ where $\vu*{x}$ lies along the $a$-axis, $\vu*{z}$ along the $c$-axis and $\vu*{y}$ completes the right-handed basis set of vectors. We define the spin rotation plane using two vectors $\vu*{u}$ and $\vu*{v}$, where we take $\vu*{u}\equiv\vu*{x}$ in the following. In order to account for a tilt of the spin rotation plane we introduce $\theta$ as the tilt angle about $\vu*{u}$ such that  $\vu*{v}=[0,-\sin\theta,\cos\theta]$. We note that in our analysis,  $\vu*{u}$ could take any direction in the $(ab)$-plane, and the definition of the tilt angle $\theta$ can be generalized. By definition, any two spins $\vb*{S}_i$ and $\vb*{S}_j$, lie within the $uv$-plane, so their cross product must lie along $\pm\vu*{n}=\vu*{u}\times\vu*{v}=[0,-\cos\theta,-\sin\theta]$ (\cref{fig:basis}).  Note that when $\theta =0$, the spins rotate in a plane containing $\vu*{z}$ and we obtain a proper cycloid [Fig. \ref{fig:MSO_struc}(d)].  When $\theta=90^\circ$, the spins rotate in a plane perpendicular to $\vu*{z}$ defining a proper helix, as reported in \bnfs. Intermediate values of $\theta$ give a generic helicoidal structure that can be decomposed into an admixture of helical and cycloidal parts.

The spin configuration within a basal triangle of Mn$^{2+}$ ions is described by the classical vector chirality ${\vb*{V}\ut{T}=\frac{1}{3}(\vb*{S}_1\times\vb*{S}_2+\vb*{S}_2\times\vb*{S}_3+\vb*{S}_3\times\vb*{S}_1)}$ where the indices are given by right hand rule around the axial vector $\vu*{z}$ defined as parallel to the positive $c$-axis. Similarly, a vector chirality $\vb*{V}\ut{C}=\vb*{S}_{\alpha}\times\vb*{S}_{\beta}$ can be introduced to describe the rotation of the spins along the $c$-axis, relatively to the polar vector $\vb*{r}_{\alpha\beta}$ where $\alpha$ and $\beta$ refer to two neighboring layers along the $c$-axis. We can now redefine the axial vector $\vb*{A}$ and the polar vector $\vb*{P}\ut{m}$ used to characterize the cycloidal magnetic structure of \mnsb\ in Ref.~\citenum{johnson2013111} as

\begin{equation}\label{eq:coschira}
\left\{
\begin{aligned}
& \vb*{A}=\vu*{z}\times\vb*{V}\ut{T}=[\ett\cos{\theta},0,0] \\
& \vb*{P}\ut{m}=\vb*{r}_{\alpha\beta}\times\vb*{V\ut{C}}=[\etc\cos{\theta},0,0]\\\end{aligned} \right.
\end{equation}

\noindent where \ett\ and \etc\ are T-even P-even and T-even P-odd parameters associated with the magnetic configuration within the $(ab)$-plane triangular motifs and on propagation along the $c$-axis, respectively. Importantly, both parameters are conserved upon rotation by $\theta$. We can similarly redefine the triangular chirality $\et$ and spin helicity $\eh$ used to characterize the helical magnetic structure of \bnfs\ in Ref. \citenum{marty2008101} as

\begin{equation}\label{eq:sinchira}
\left\{
\begin{aligned}
& \et = \vu*{z}\cdot\vb*{V}\ut{T}=-\ett\sin{\theta}\\
& \eh = \vb*{r}_{\alpha\beta}\cdot\vb*{V\ut{C}}=-\etc\sin{\theta}\\
\end{aligned} \right.
\end{equation}
These expressions allow us to use \ett, \etc, and $\theta$ to parametrize a generic helicoidal magnetic structure. The vector quantities of \cref{eq:coschira} capture the cycloidal component projected into the ($ac$)-plane, and the scalar quantities of \cref{eq:sinchira} capture the helical part projected into the ($ab$)-plane. We note that the helical part is odd in $\theta$, while the cycloidal part is even.

\subsection{Magnetic structure description}\label{sec:magstruc}

Considering the two perpendicular unit vectors $\vu*{u}$ and $\vu*{v}$ that define the spin rotation plane, we can describe the magnetic moment for a Mn atom at site $j=(1,2,3)$ on a given triangular motif, in layer $\alpha$ (along the $c$-axis), and with an angle $\phi_{\alpha j}$:\cite{kinoshita2016117}

\begin{equation}\label{eq:moments}
\left\{
\begin{aligned}
& \vb*{\mu}_{\alpha j}=M_{u}\cos{\phi_{\alpha j}}\vu*{u}+M_{v}\sin{\phi_{\alpha j}}\vu*{v} \\
& \phi_{\alpha j} = 2\pi\etc k\ut{z}\alpha+\ett (j-1)\frac{2\pi}{3} \end{aligned} \right.
\end{equation}

\noindent $M_{u}$ and $M_{v}$ describe the shape of the ellipse (circular for $M_{u}=M_{v}$), $k\ut{z}$ is the vertical component of the propagation vector $\vb*{k}=(0,0,k\ut{z})$. \etc\ and \ett\ describe the sense of rotation of the spins respectively along the positive $c$-axis, and within a Mn$^{2+}$ triangle, following the definitions above.

The first magnetic structure proposed in Ref.~\citenum{johnson2013111} has $\vu*{u}$ lying along the crystallographic $a$-axis, and $\vu*{v}$ along the $c$-axis. This magnetic structure preserves the twofold symmetry (magnetic space group B21'). This is not the case in the model proposed in Ref.~\citenum{kinoshita2016117}, with $\vu*{u} \parallel [1\bar{1}0]$, which lowers the symmetry of the magnetic space group to P11' owing to the breaking of the twofold symmetry. However, the tilting of $\vu*{v}$ from the $c$-axis reported in Ref.~\citenum{kinoshita2016117}, by an angle $\theta$, is also allowed in the B21' space group as long as the twofold symmetry is preserved.  Both models consider the presence of threefold domains, and for each of them, the magnetic moments in \cref{eq:moments} are transformed by rotating $\vu*{u}$ and $\vu*{v}$ by $\SI{120}{\degree}$ around the $c$-axis.

\subsection{Invariant from Heisenberg interactions}

If we consider a Heisenberg Hamiltonian with seven SSE pathways,\cite{johnson2013111} the classical mean-field energy can be derived as a function of the propagation vector $k\ut{z}>0$: 

\begin{align}\label{eq:energy}
E_0(k)&=-\frac{1}{2}(J_1+2J_2)+J_4\cos{(2\pi\etc k\ut{z})} \nonumber\\
&+J\ut{R}\cos{(2\pi\etc k\ut{z}+\ett\frac{2\pi}{3})}+J\ut{L}\cos{(2\pi\etc k\ut{z}-\ett\frac{2\pi}{3})}
\end{align}

\noindent where $J\ut{R}=J_3+2J_6$ sums the right-handed interactions and $J\ut{L}=J_5+2J_7$ sums the left-handed interactions. Minimizing \cref{eq:energy} with respect to the propagation vector gives for the ground state:

\begin{equation}\label{eq:tan}
\tan\left({2\pi\etc k\ut{z}}\right)=\frac{\ett\sqrt{3}(J\ut{R}-J\ut{L})}{J_\text{R}+J_\text{L}-2J_4}
\end{equation}

As left-handed and right-handed exchange paths are switched between the enantiomorphs, the quantity $J\ut{R}-J\ut{L}$ changes sign upon inversion symmetry. Thus taking the DFT values for the exchange constants from Ref.~\citenum{johnson2013111}, a sign analysis of \cref{eq:tan} gives the invariant:

\begin{equation}\label{eq:invariant}
\sigma\etc\ett=+1
\end{equation}

\noindent where $\sigma=+1$ for a left-handed crystal structure (L), and $\sigma=-1$ for a right-handed crystal structure (R). This is similar to iron langasite, where the structural chirality is linked to the pair of magnetic chiralities readily obtained by substituting $\theta = 90^\circ$ into  \cref{eq:sinchira}.\cite{marty2008101}

\section{Results and discussion}\label{sec:results}

Having outlined the experimental neutron diffraction techniques and the definitions relevant for the discussion of MnSb$_{2}$O$_{6}$, we now present the experimental results.  We first discuss the nuclear and then the low temperature magnetic structure. 

\subsection{Nuclear structure}

\subsubsection{Single crystal neutron diffraction}

For a given Bragg reflection \Q, the inversion twin will scatter with a nuclear structure factor $F\ut{N}(-\Q)$. In absence of resonant scattering, Friedel's law is valid, and both twins will scatter the same nuclear intensity $\propto |F\ut{N}(\Q)|^2$. Inversion twins are thus indistinguishable by unpolarized neutrons. On the other hand, twofold twins reveal different nuclear structure factors depending on the $(hkl)$ indices so their domain population can be refined using unpolarized neutrons if the appropriate Bragg reflections are measured. We collected intensities from 430 nuclear reflections at \SI{50}{K} on the four-circle diffractometer D9. Rocking scans show nicely resolved Bragg peaks, with a full width at half‐maximum $\sim 0.4^\circ$ in $\omega$. The data were refined using \textsc{Fullprof}.~\cite{rodriguez-carvajal1993192} The parameters scale, extinction, atomic positions, displacements, as well as domain population for twofold twins were refined, showing that our single crystal has no twofold twins as one nuclear intensity domain was refined to a population of 0.991(3). Our refinement results (detailed in \cref{tab:nuc_table}) agree with the known crystal structure previously studied by neutron powder diffraction at room temperature.\cite{reimers1989} 

\begin{figure*}[]
    \begin{center}
    \includegraphics[scale=1]{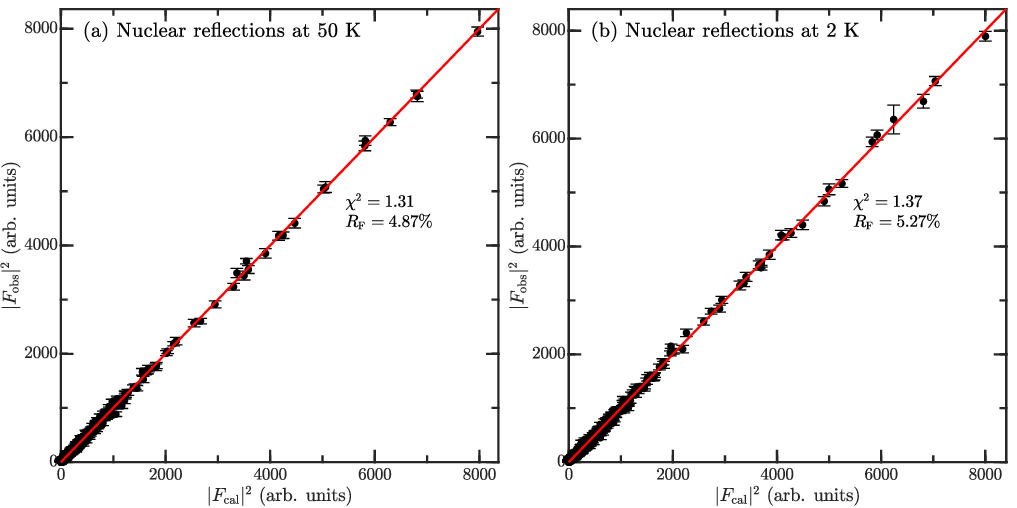}
    \end{center}
    \caption{Observed versus calculated intensities in $P$321 space group for nuclear reflections measured at (a) \SI{50}{K}, (b) \SI{2}{K}.}
    \label{fig:Fobs_Fcal}
\end{figure*}

As the threefold symmetry from paramagnetic $P$321 space group is broken by the cycloidal structure in the magnetic phase,\cite{johnson2013111} this could relate to a symmetry lowering of the nuclear space group below $T\ut{N}$. To investigate the possibility of a structural distortion coinciding with $T\ut{N}$, a separate set of 318 Bragg reflections was measured at \SI{2}{K} (below $\tn\approx\SI{12}{K}$), leading to 75 inequivalent groups of reflections. If the crystal symmetry is reduced, the equivalent reflections in $P$321 should no longer be equivalent within each group of reflections. For example, reflections $(h,k,l)$, $(k,-h-k,l)$ and $(-h-k,h,l)$ are related by threefold symmetry along the $c$-axis and are thus equivalent in $P$321. In the case where the threefold symmetry is broken, these three kind of reflections are no more equivalent. In addition, three structural domains rotated by \SI{120}{\degree} are expected. If these threefold domains are exactly equi-populated, the intensities scattered from each domain will average out, making them impossible to be distinguished from a single threefold symmetric domain. Else, the intensities of reflections within a group of $P$321-equivalent reflections will differ. The internal $R$-factor is $R\ut{int}=4.1\%$ for the data reduction in $P$321 symmetry, which indicates that the differences of intensities for $P$321-equivalent reflections are not measurable given our setup. In addition, the data was refined including the threefold domains in $P$1 symmetry, but this did not significantly improve the refinement. 
At the end 845 nuclear reflections were measured at \SI{2}{K}, and were well refined in $P$321 space group as shown in \cref{fig:Fobs_Fcal}(b), in comparison to the \SI{50}{K} refinement in \cref{fig:Fobs_Fcal}(a). From this, there is no significant evidence of breaking of $P$321 symmetry below N\'
{e}el temperature. Detailed refinement results for both temperatures are listed in \cref{tab:nuc_table}.

\renewcommand{\arraystretch}{1.05} 
\begin{table*}[]
\caption{Structural parameters of \mnsb\ single crystal measured on D9, refined with \textsc{Fullprof}\cite{rodriguez-carvajal1993192}  within nuclear space group \textit{P}321 (No.~150)}
\label{tab:nuc_table}
\begin{tabularx}{\textwidth}{XXXXXccX}
\hline \hline

\multicolumn{7}{l}{$T=\SI{50}{K}$}{Measured, independent, observations with equivalent reflections: 430, 406, 44}\\
\multicolumn{7}{l}{$R\ut{int}=8.78\%$ \quad $R\ut{F}=4.87\%$ \quad $R\ut{Bragg}=4.69\%$ \quad $\chi^2=1.31$} \\ 
\multicolumn{7}{l}{$a=b = 8.7835(81)$ \AA \quad $c$ = 4.7238(58) \AA} \\ \hline

Atoms & Wyckoff  &  $x$   &  $y$  &  $z$  &  $B\ut{iso}$ (\AA$^2$) & Occ. \\ \hline
Mn  &   3$e$    &  0.6319(3)  &  0.0000  &  0.0000  &  0.19(3) & 1 \\ 
Sb1 &   1$a$    &  0.0000  &  0.0000  &  0.0000  &  0.06(3) & 1 \\ 
Sb2 &   2$d$    &  0.3333  &  0.6667  &  0.5059(4)  &  0.04(3) & 1 \\ 
Sb3 &   3$f$    &  0.3050(3)  &  0.0000  &  0.5000  &  0.09(2) & 1 \\
O1  &   6$g$    &  0.1046(3)  &  0.8917(3)  &  0.7626(2)  &  0.24(2) & 1 \\ 
O2  &   6$g$    &  0.4711(2)  &  0.5891(2)  &  0.7286(2)  &  0.19(2) & 1 \\ 
O3  &   6$g$    &  0.2258(3)  &  0.7804(3)  &  0.2805(2)  &  0.16(2) & 1 \\  

\end{tabularx}
\begin{tabularx}{\textwidth}{XXXXXccX}
\hline  
\multicolumn{7}{l}{$T=\SI{2}{K}$}{Measured, independent, observations with equivalent reflections: 845, 529, 423}\\
\multicolumn{7}{l}{$R\ut{int}=4.09\%$ \quad $R\ut{F}=5.27\%$ \quad $R\ut{Bragg}=5.40\%$ \quad $\chi^2=1.37$} \\ 
\multicolumn{7}{l}{$a=b = 8.7907(19)$ \AA \quad $c$ = 4.7176(10) \AA} \\ \hline

Atoms & Wyckoff  &  $x$   &  $y$  &  $z$  &  $B\ut{iso}$ (\AA$^2$) & Occ. \\ \hline
Mn  &   3$e$    &  0.6329(3)  &  0.0000  &  0.0000  &  0.30(3) & 1 \\ 
Sb1 &   1$a$    &  0.0000  &  0.0000  &  0.0000  &  0.16(3) & 1 \\ 
Sb2 &   2$d$    &  0.3333  &  0.6667  &  0.5061(5)  &  0.09(3) & 1 \\ 
Sb3 &   3$f$    &  0.3050(2)  &  0.0000  &  0.5000  &  0.09(2) & 1 \\
O1  &   6$g$    &  0.1047(2)  &  0.8920(3)  &  0.7628(2)  &  0.27(1) & 1 \\ 
O2  &   6$g$    &  0.4710(2)  &  0.5889(2)  &  0.7285(2)  &  0.25(2) & 1 \\ 
O3  &   6$g$    &  0.2253(3)  &  0.7799(2)  &  0.2804(2)  &  0.23(2) & 1 \\  
\hline\hline
\end{tabularx}

\end{table*}

\subsubsection{Schwinger scattering}\label{sec:schwinger}

To characterize the chiral domains, Schwinger scattering was measured on D3 on nine Bragg reflections at $T=\SI{3}{K}$ on the same single crystal characterized on D9, for which only two out of four possible twins were measured to be present as explained above. Absolute indexation was determined on D9 by comparing the nuclear intensities of Bragg reflections. This was not done on D3 (as only flipping ratios were measured), so the reflections can be indexed with a twofold rotation between D3 and D9 experiments. Thus, either \{L$(hkl)$, R$(\bar{h}\bar{k}\bar{l})$\}, or \{L$(\bar{h}\bar{k}l)$, R$(hk\bar{l})$\} are the twins present (with the indexation from D3 experiment).

\begin{figure}[h]
    \begin{center}
    \includegraphics{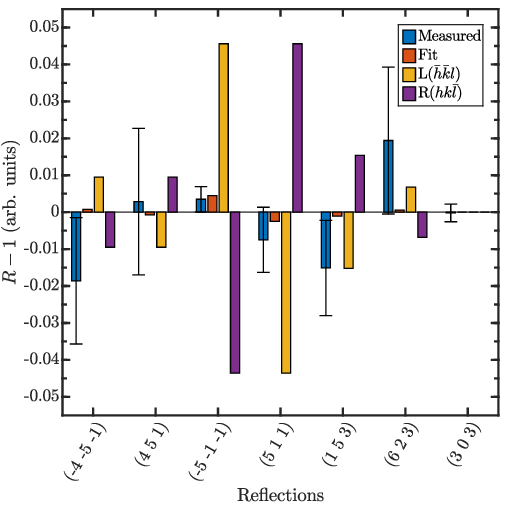}
    \end{center}
    \caption{Measured flipping ratios are fitted to a linear combination of the theoretical flipping ratios for two structural twins.}
    \label{fig:schwinger}
\end{figure}

The experimental flipping ratios are then fitted to a linear combination of the theoretical ones (calculated with the atomic positions from D9 data refinement at \SI{2}{K}), as shown in \cref{fig:schwinger}. The best fit is obtained considering the twins L$(\bar{h}\bar{k}l)$ and R$(hk\bar{l})$, giving 0.54(2) of left-handed structural domain, and 0.46(2) of right-handed domain. The error bars are quite large in this experiment, but the flipping ratios being close but different than 1 within uncertainties indicate that there is a mixture of chiral inversion twins in the crystal. For an enantiopure, the flipping ratios should be close to one set of predicted flipping ratios, which shows much more pronounced asymmetries as exemplified by the (511) and (153) reflections. The results are thus different from enantiopure \bnfs\ single crystals which were previously studied.\cite{marty2008101,stock201183,qureshi2020102}

\subsubsection{Transmission Polarized Optical Microscopy}

Chiral structural domains in a single crystal can also be measured with a polarized optical microscope. Due to the optical activity in chiral compounds, the polarization plane of a linearly polarized light is rotated after traveling through the sample.\cite{jerphagnon197665} The sense of rotation depends on the handedness of the considered domains, which can be distinguished by observing the transmitted light through an analyzer.\cite{wang20153,prosnikov2019100}

\begin{figure}[h]
    \begin{center}
    \includegraphics{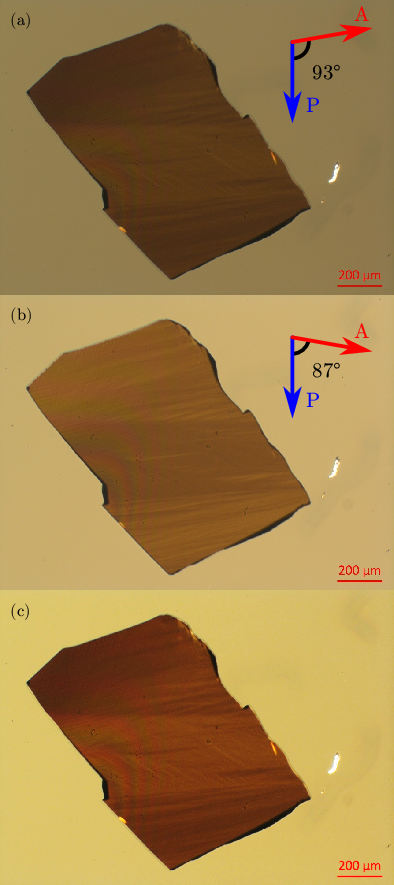}
    \end{center}
    \caption{Transmission polarized optical microscopy images of a single crystal of \mnsb : for different angles between the polarizer (P) and analyzer (A) in (a) and (b). (c) Images substracted, showing the chiral domains in the sample. }
    \label{fig:TPOM}
\end{figure}

A different sample of \mnsb, synthesized following the same procedure described in \cref{ssec:preparation}, was observed under a transmission polarized optical microscope. The directions of the polarizer and analyzer are shown in blue and red in \cref{fig:TPOM}(a)-(b), forming an angle $\theta=90\pm3^\circ$. These images show several domains with opposite chirality. The constrast between neighboring domains is reverted by rotating the analyzer from $\theta=93^\circ$ to $\theta=87^\circ$ because the polarization plane of the transmitted light is rotated in the opposite sense for opposite chirality domains in the sample. \cref{fig:TPOM}(c) shows the difference of intensity between \cref{fig:TPOM}(a) and \cref{fig:TPOM}(b), clearly revealing the chiral areas in the single crystal.

Given the same chemical synthesis, our other single crystals, including the one studied under neutron diffraction, are likely to have a similar behavior. They are expected to be a mixture of chiral structural domains, which is consistent with our Schwinger scattering analysis described above.

\subsubsection{Magneto-structural effects}

\begin{figure}[h]
    \begin{center}
    \includegraphics{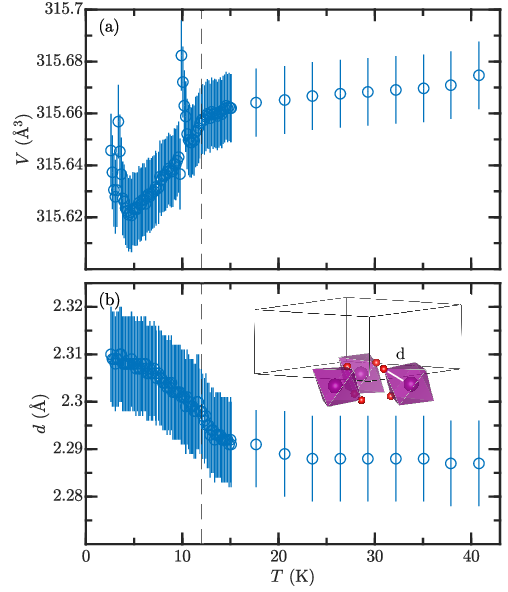}
    \end{center}
    \caption{Refinement results from D20. Temperature dependence of: (a) the unit cell volume, (b) the bond length between Mn (in purple) and symmetry equivalent O1 atoms (in red). $\tn\approx\SI{12}{K}$ is shown in dashed gray lines.}
    \label{fig:D20_struc}
\end{figure}

Neutron powder diffraction was performed on D20 from \SI{2.5}{K} to \SI{89.5}{K}. The nuclear structure was refined sequentially as a function of temperature using \textsc{Fullprof}.\cite{rodriguez-carvajal1993192} While no symmetry breaking of the $P$321 paramagnetic space group was evidenced by our studies, as discussed above, structural changes induced by the phase transition are visible from the powder diffraction data refinement. \cref{fig:D20_struc}(a) shows the refined volume of the unit cell as a function of temperature. The volume decreases sharply under $\tn\approx\SI{12}{K}$, demonstrating a deviation from the linear thermal expansion of the unit cell upon magnetic ordering. Actually, this results from the contraction of both $a$ and $c$ lattice constants. Similarly, changes in bond distances are caused by magneto-elastic effects, as shown in \cref{fig:D20_struc}(b) for the distance between Mn atom (in purple) and symmetry equivalent O1 atoms (in red). We note that the unit cell volume shows some anomalies in \cref{fig:D20_struc}(a) around $\SI{3}{K}$ and $\SI{10}{K}$. We have over-plotted the different diffraction patterns and could not observe any shift in the peaks positions. We think that these jumps are numerical artifacts rather than real lattice parameters shifts.

\subsection{Magnetic structure}

\subsubsection{Order parameter}

\begin{figure}[h]
    \begin{center}
    \includegraphics{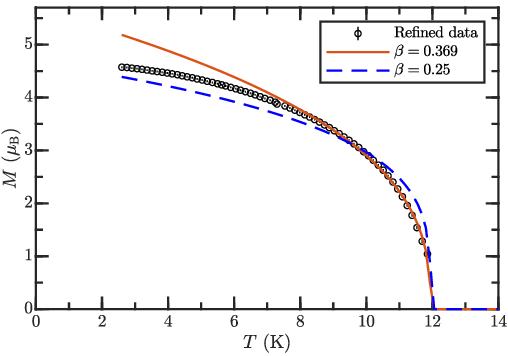}
    \end{center}
    \caption{Refined magnetic moments from D20 as a function of temperature, fitted to a power law $\propto (T-\tn)^\beta$ with the critical exponent $\beta$ fixed for 3D Heisenberg model (solid red curve) and for 2D $XY$ model (dashed blue curve).}
    \label{fig:D20moments}
\end{figure}

Neutron powder diffraction is not sensitive to the direction of the magnetic moments in the ($ab$)-plane, and neither to the magnetic chiralities. Yet the magnitude of the magnetic moments can be refined from D20 powder diffraction data, as a function of temperature. The cycloid was constrained to be circular ($M_{u}=M_{v}$) and the refined moments are shown in \cref{fig:D20moments}. The data in the critical region ($\SI{8}{K}<T<\SI{12}{K}$) are fitted to a power law $\propto (T-\tn)^\beta$, with the critical exponent fixed to $\beta=0.369$ (solid red curve) as expected for the nonfrustrated 3D Heisenberg model,\cite{campostrini200265} and to $\beta=0.25$ (dashed blue curve) measured for iron langasite\cite{stock201183} and $XY$-like stacked-triangular magnets.\cite{kawamura198863} The critical behavior near $\tn=\SI{11.94(1)}{K}$ is in agreement with the 3D Heisenberg model as suggested previously in Ref.~\citenum{reimers1989}. Therefore \mnsb\ does not have the same universality class as iron langasite and other layered-triangular magnets.

\subsubsection{Unpolarized single crystal diffraction}

From the invariant derived in \cref{eq:invariant} ($\sigma \eta_{C} \eta_{T}=1$), a given structural chirality $\sigma$ is compatible with two pairs of magnetic configurations $(\etc,\ett)$. We can label the structural and magnetic configurations as $\sigma(\etc,\ett)$, which gives four possibilities L$(1,1)$, L$(-1,-1)$, R$(-1,1)$, R$(1,-1)$. L$(1,1)$ and L$(-1,-1)$ configurations lead to the same magnetic intensities, and R$(-1,1)$, R$(1,-1)$ are the respective configurations of their inversion twins, as \etc\ is P-odd and \ett\ P-even from \cref{eq:coschira}. Magnetic intensities of inversion twins satisfy Friedel's law, so the four configurations are undistinguishable by unpolarized neutrons. However, as mentioned above, twofold structural twins can exist in the $P$321 space group, leading to a different set of nuclear and magnetic intensities (see \cref{tab:twin}).

\renewcommand{\arraystretch}{1.1} 
\begin{table}[h]
    \centering
    \begin{tabular}{ccccccc}    
    \hline
    \hline
         Twin & $\sigma=\etc\ett$ & $|F\ut{N}|^2$ & $|\Mperp|^2$ & $R$ \\ \hline
         \color{red}L$(hkl)$  & \color{red}$+1$ & \color{red}$N_1$ &\color{red} $M_1$ & \color{red}$R_1$\\ 
         L$(\bar{h}\bar{k}l)$  & $+1$ & $N_2$ & $M_2$ & $R_2$\\ 
         \color{red}R$(\bar{h}\bar{k}\bar{l})$  & \color{red}$-1$& \color{red}$N_1$ & \color{red}$M_1$ & \color{red}$R_3$\\ 
         R$(hk\bar{l})$ & $-1$ & $N_2$ & $M_2$ & $R_4$ \\ \hline  \hline  
    \end{tabular}
    \caption{Summary of the possible twins and their sensitivity to nuclear ($|F\ut{N}|^2$) and magnetic diffraction ($|\Mperp|^2$), and Schwinger scattering (flipping ratio $R$). Different subscripts denote different values. The twins present in our single crystal are highlighted in red.}
    \label{tab:twin}
\end{table}

In previous studies, unpolarized neutron single crystal diffraction data were refined with a mixture of two sets of calculated magnetic intensities, attributed to two chiral structural domains. In light of the present study, one should actually assign these two sets of intensities to at least two twofold domains, with a potential further mixture of chiral domains to which the experiment was not sensitive. In Ref.~\citenum{johnson2013111}, the single crystal neutron diffraction magnetic refinement shows a 0.8(1):0.2(2) domain fraction of the calculated intensities, which corresponds to a fraction 0.8 of twins \{L$(hkl)$, R$(\bar{h}\bar{k}\bar{l})$\}, and 0.2 of twins \{L$(\bar{h}\bar{k}l)$, R$(hk\bar{l})$\}. In absence of a method (Schwinger scattering or anomalous x-ray scattering) sensitive to the inversion twins, one cannot conclude on the population of all four domains. A similar issue arose in Ref.~\citenum{kinoshita2016117}, where only one set of magnetic intensities was found and attributed to an enantiopure crystal, but could actually include a mixture of a twin and its chiral inversion twin.

The same single crystal characterized on D3 and D9 was measured on D10. The magnetic structure was refined using \textsc{Mag2Pol}\cite{qureshi201952} (cross-checked with \textsc{Fullprof},\cite{rodriguez-carvajal1993192} giving similar results), with 256 magnetic reflections collected at \SI{2}{K}. The scale and extinction parameters are refined using 145 nuclear reflections (40 inequivalent, giving $R\ut{F}=4.88\%$). A single domain in terms of magnetic intensities was found, meaning the absence of twofold structural twins and confirming our results from D9. These intensities are consistent with two twins related by inversion symmetry, shown in red in \cref{tab:twin}, which can be distinguished by Schwinger scattering (see \cref{sec:schwinger}). Extinction parameters can be significantly different for nuclear and magnetic reflections, due to multiple magnetic domains having smaller sizes than the structural domains\cite{qureshi200979}. This is the case from our refinement, where the extinction parameters refined with the magnetic intensities are found smaller than the one refined with the nuclear intensities. To keep a consistent comparison between the magnetic structure models, the extinction parameters were set to zero for the magnetic refinement described below.

\begin{table}[h]
    \centering
    \begin{tabular}{ccccccccc}    
    \hline\hline
         Name &$\vu*{u}$ & $M_u$ & $M_v$ & $\theta\,(^\circ)$ & $p_1$ & $p_2$ & $p_3$ & $R\ut{F}$ (\%) \\  \hline
        A &$\vu*{a}$ & $4.5(1)$ & $4.7(1)$ & $0$ & $1$ & $0$ & $0$ & $19.26$\\ 
        B &$\vu*{a}$ & $5.7(1)$ & $3.7(1)$ & $0$ & $0.40$ & $0.20(3)$ & $0.40(3)$ & $15.29$\\ 
        C &$\vu*{a}$ & $5.6(3)$ & $3.8(3)$ & $9(28)$ & $0.40$ & $0.20(5)$ & $0.40(3)$ & $15.31$\\ 
        D &$[1\bar{1}0]$ & $5.7(1)$ & $3.7(1)$ & $0$ & $0.27$ & $0.27(3)$ & $0.46(3)$ & $15.29$\\ 
        E &$[1\bar{1}0]$ & $5.9(2)$ & $3.8(3)$ & $15(14)$ & $0.28$ & $0.25(4)$ & $0.47(4)$ & $15.26$ \\ \hline\hline
    \end{tabular}
    \caption{Refined parameters obtained for non-tilted and tilted cycloidal models.}
    \label{tab:D10}
\end{table}

\begin{figure}[h]
    \begin{center}
    \includegraphics{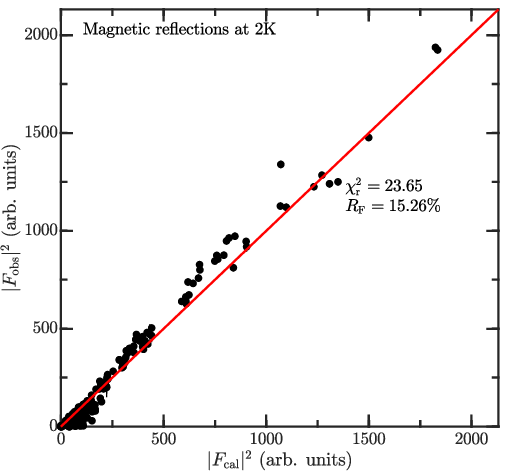}
    \end{center}
    \caption{Observed versus calculated intensities for magnetic reflections measured at \SI{2}{K}.}
    \label{fig:D10_fit}
\end{figure}

The refinement results using different magnetic structure models labeled from A to E are listed in \cref{tab:D10}. While including the threefold domains (A$\rightarrow$B) with populations $p_1$, $p_2$, and $p_3$ improves the goodness of fit, there is no observable difference between models with the in-plane main axis $\vu*{u}$ of the cycloid along the $a$-axis and along $[1\bar{1}0]$ (B$\rightarrow$D). Similarly, allowing a tilt around the $a$-axis (B$\rightarrow$C), and around $[1\bar{1}0]$ (D$\rightarrow$E) does not significantly improve the fit. This is because the in-plane direction $\vu*{u}$ of the spin rotation plane, and the tilt angle $\theta$ are correlated with the magnetic domain fractions, which makes no much difference in terms of goodness of fit between models B, C, D and E. Our best fit with the model considered in Ref.~\citenum{kinoshita2016117} is obtained with a tilt angle $\theta=+15(14)^\circ$ (\cref{fig:D10_fit}), compared to previously found $\theta=18(5)^\circ$. However, two equi-populated tilt domains with $\theta=\pm 18(5)$ were considered in Ref.~\citenum{kinoshita2016117} while in our refinement, a single tilt domain $\theta>0$ was more consistent. Based on our single crystal diffraction data, we however do not observe a significant improvement in the resulting fit with inclusion of a tilt in the magnetic structure.

\subsubsection{Spherical neutron polarimetry}

SNP was performed on D3, using the same experimental setup as for the Schwinger experiment with the exception of a $^3$He spin filter necessary for the polarization analysis of the final neutron beam. The full polarization matrices of five magnetic Bragg reflections were measured at $T=\SI{3}{K}$. CRYOPAD\cite{tasset1999267-268a} is used to protect the sample from any external magnetic fields, and to select independently the initial and final polarization directions of the neutrons. In the case of \mnsb, SNP is sensitive to the threefold magnetic domains and the cycloidal parameter $\etc$, while the triangular parameter $\ett$ can not be distinguished. The measured polarization matrices were fitted using \textsc{Mag2Pol}\cite{qureshi201952} to a linear combination of the possible polarization matrices as $P\ut{meas}=\sum_{i} \alpha_i P_i$ with $\alpha_i$, $P_i$, the population and polarization matrix of the $i$-th magnetic domain. 

The magnetic moments were first refined in the $ac$-plane ($\vu*{u}=\vu*{a}$ and $\vu*{v}=\vu*{c}$). In the absence of a nuclear contribution to the scattered intensity, SNP is not sensitive to the size of the magnetic moments. Therefore, since in this experiment purely magnetic satellites are investigated, only the ratio $e=M_v/M_u$, known as the ellipticity, can be deduced. Considering the model proposed in Ref.~\citenum{johnson2013111}, threefold and $\etc=\pm 1$ domain populations are refined, leading to six polarization matrices to consider. The refinement results for this model are shown in \cref{tab:SNP_notilt}. The cycloids are found elliptical along the basal direction with $e=0.92(1)$ and $\chi^2\ut{r}=7.14$. The population for the third threefold domain with $\etc=+1$ was fixed to 0 in order to avoid fit divergence and unphysical results. 

\begin{table}[h]
    \centering
    \begin{minipage}[t]{0.65\linewidth}
        \begin{tabular}{ccccc}
          \hline\hline
          \etc & $1$ & $3^+_z$ & $3^-_z$ & Sum \\ \hline
          +1 & 0.20(1) & 0.20(1) & 0 & 0.40(2) \\ 
          $^-$1 & 0.09(1) & 0.09(1) & 0.42(2) & 0.60(2) \\  \hline
          Sum & 0.29(1) & 0.28(1) & 0.42(2) & 1 \\ \hline\hline
        \end{tabular}
    \end{minipage}
    \begin{minipage}[t]{0.3\linewidth}
    \begin{tabular}{cc}
        \hline\hline
        $\chi^2\ut{r}$ & 7.14 \\ \hline
        $e$& 0.92(1) \\ \hline\hline
        \end{tabular}
    \end{minipage}
    \caption{Refined parameters for the non-tilted cycloid model.} 
    \label{tab:SNP_notilt}
\end{table}

The SNP data were then fitted to the tilted cycloid model proposed in Ref.~\citenum{kinoshita2016117}.  In this case, the positive and negative tilt of the angle $\theta$ have to be taken account because it changes the rotation plane and leads to different polarization matrices. This doubles the number of polarization matrices to include, resulting in 12 domain populations to refine (threefold $\times\{\etc=\pm 1\}\times\pm\theta$). The vectors $\vu*{u}$ and $\vu*{v}$ of each of these 12 magnetic domains are related by symmetry operators and the absolute values of $M_u$  are constrained to be the same for each magnetic domain (the same for $M_v$), so that each magnetic domain keeps the same magnetic moment size. This also constrains the absolute value of the tilt angle to be the same for $\theta>0$ and $\theta<0$ domains. The results are shown in \cref{tab:SNP_tilt}. Again, the domains returning unphysical values in a first refinement step were fixed to zero in the following. The positive tilt domains are predominant, with a population of 0.89(4), giving $\theta=14(7)^\circ$ which is consistent with the best fit from the D10 data. However, this tilted model only slightly improves the goodness of fit to $\chi^2\ut{r}=6.68$. 

\begin{table}[h]
    \centering
    \begin{minipage}[t]{0.73\linewidth}
        \begin{tabular}{cccccc}
          \hline\hline
          \etc & $\theta$ & $1$ & $3^+_z$ & $3^-_z$ & Sum\\\hline
          $-1$ & + & 0.08(1) & 0.32(3) & 0 & 0.40(3)\\ 
          +1 & - & 0.01(2) & 0.07(2) & 0.04(2) & 0.12(3)\\ 
          $-$1 & - & 0 & 0 & 0 & 0\\ 
          +1 & + & 0.14(1) & 0 & 0.35(2) & 0.49(2)\\   \hline 
          Sum & & 0.23(2) & 0.39(4) & 0.39(2) & 1\\ \hline\hline
        \end{tabular}
    \end{minipage}
    \begin{minipage}[t]{0.2\linewidth}
    \begin{tabular}{cc}
        \hline\hline
        $\chi^2\ut{r}$ & 6.68 \\ \hline
        $e$& 0.96(8) \\ \hline
        $\theta$ & $14(7)^\circ$\\ \hline\hline
        \end{tabular}
    \end{minipage}
    \caption{Refined parameters for the tilted cycloid model.}
    \label{tab:SNP_tilt}
\end{table}

Our diffraction study of the magnetic structure of \mnsb\ evidences a mixture of threefold magnetic domains and magnetic polarities. In the absence of a substantial improvement in $R$-factors on inclusion of the model with in-plane moments along $[1\bar{1}0]$, we propose that the model with moments along $\vu*{a}$ is the ground state because it has a higher symmetry (not breaking twofold symmetry). The tilt is still allowed by symmetry, as pointed out in \cref{sec:magstruc}. Thus the possibility of a tilted cycloidal structure is not ruled out by symmetry considerations our experiments.
In \cref{sec:mf} we discuss the appearance of a macroscopic electric polarization reported in Ref.~\citenum{kinoshita2016117} and propose a different mechanism without invoking the need of a tilted cycloid ground state. 

\subsection{Magnetic field dependence}

Before discussing the electric polarization we finally investigate the magnetic field response of the magnetic structure in MnSb$_{2}$O$_{6}$ owing to its importance in any domain switching.  Magnetic phase transitions induced by low magnetic fields (below $\SI{2}{T}$) were observed previously in \mnsb\ bulk magnetization measurements.\cite{werner201694,kinoshita2016117} This was explained by a very small anisotropy stabilizing the cycloidal magnetic ground state, which can be easily overcome by applying a magnetic field, changing the spin structure to another state. In order to complement these macroscopic measurements, neutron diffraction was performed on RITA-2 using a horizontal magnetic field, on a single crystal of \mnsb, aligned in the $(H,0,L)$ scattering plane such that the magnetic field could be aligned either along the $c$ or $a$-axes. A single high intensity magnetic peak, $\Q=(1,0,1)-\vb*{k}$ was scanned over a range of temperatures (between 1.75 and 11.5 K) and magnetic fields (between 0 and 5 T), applied parallel and perpendicular to the $c$-axis. Unpolarized neutrons are sensitive to the magnetic moments perpendicular to the scattering vector $\Q$, so a change of the measured intensity can be a direct proof of a change in the magnetic structure.

\begin{figure}[h]
    \begin{center}
    \includegraphics{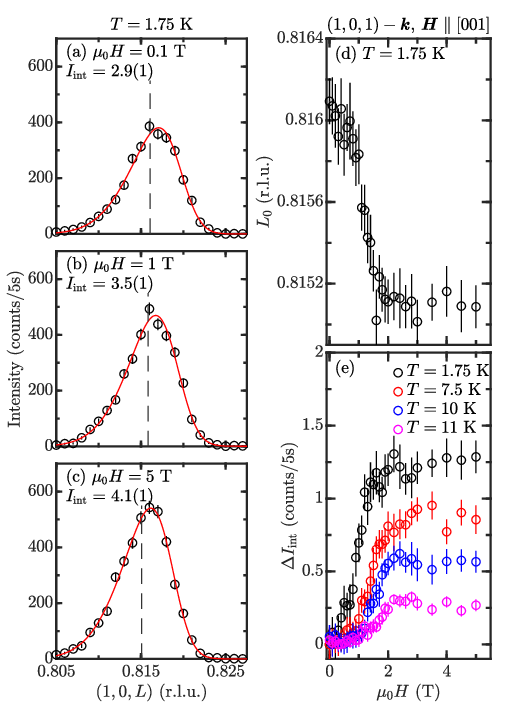}
    \end{center}
    \caption{The magnetic field is applied along the $c$-axis. (a)-(c) Scans at $T=\SI{1.75}{K}$ along the $(00L)$ direction for different values of magnetic field. The mean position of the skewed Gaussian fits are shown in dashed lines and depend on the applied field. (d) Summary of the field dependence of the propagation vector at $T=\SI{1.75}{K}$. (e) Integrated intensities of the magnetic peak $\Q=(1,0,1)-\vb*{k}$ as a function of the magnetic field. The zero-field intensity is subtracted from each respective curve for a clearer comparison of the field-induced intensity increase.}
    \label{fig:RITA_par}
\end{figure}

\cref{fig:RITA_par} shows the results for the magnetic field applied along the $c$-axis. \cref{fig:RITA_par}(a)-(c) show reciprocal space scans along the $L$ direction of the $\Q=(1,0,1)-\vb*{k}$ magnetic peak at $T=\SI{1.75}{K}$. The intensities are fitted to a skewed Gaussian: 

\begin{equation}\label{eq:skewed}
\begin{aligned}
I(L) \propto \left\{1+\erf{\left[\frac{\gamma(L-L_0)}{\sigma\sqrt{2}}\right]}\right\}\exp{-(L-L_0)^2 \over 2\sigma^2} 
\end{aligned} 
\end{equation}

\noindent where $\gamma$ is the skewness parameter, $\sigma$ and $L_0$ are the Gaussian standard deviation and center. The mean values of the skew Gaussian are shown in dashed gray lines and change with the magnetic field. A nuclear reflection $(2 0 1)$ was also monitored as a function of the magnetic field and does not present any shift along the $L$ direction. This means that the shift of the magnetic peak $\Q=(1,0,1)-\vb*{k}$ is caused by a change of the propagation $\vb*{k}$ and not of the lattice parameter $c$. This is summarized in \cref{fig:RITA_par}(d) where the propagation vector evolution can clearly be observed until a threshold magnetic field (around $\SI{2}{T}$). Magneto-elastic effects can be induced by a change in the magnetic structure as illustrated in the change in bond distances at \tn\ discussed above and shown in \cref{fig:D20_struc}. A change in the bond distances would result in a change in the strength of the exchange constants, which consequently change the propagation vector in order to minimize the ground state energy, from \cref{eq:tan}.

\begin{figure}[h]
    \begin{center}
    \includegraphics{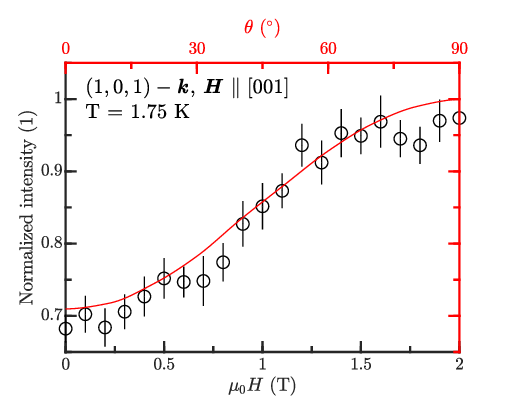}
    \end{center}
    \caption{(black points) Normalized integrated intensity of the measured magnetic peak $\Q=(1,0,1)-\vb*{k}$ as a function of the magnetic field applied parallel to the $c$-axis at $T=\SI{1.75}{K}$. (red curve) Simulated magnetic intensity as a function of the tilt angle $\theta$ of the spin rotation plane from the $c$-axis.}
    \label{fig:RITA_Ivt}
\end{figure}

In \cref{fig:RITA_par}(e), the integrated intensities are displayed as a function of the magnetic field, for different temperatures. The zero-field intensity is subtracted from each respective curve, in order to compare the data on the same scale as the magnetic intensity diminishes when the temperature increases. The integrated intensities increase with the magnetic field until a threshold value (different for each temperature) and then remain constant. For a cycloidal magnetic ground state, when no external field is applied, one main axis of the spin ellipse lies in the $(ab)$-plane, and the other one along the $c$-axis. Applying a magnetic field $\vb*{H}\parallel\vb*{c}$ is expected to flop the spin rotation plane from a cycloid to a helix, where the latter is oriented perpendicular to the magnetic field. The gradual increase of the intensity shows that the cycloid plane is continously tilted from the $c$-axis. For $T=\SI{1.75}{K}$, the observed intensities (with a magnetic field $<\SI{2}{T}$) of reflection $\Q=(1,0,1)-\vb*{k}$ match with calculated intensities for a circular helicoidal magnetic structure (with the main axis $\vu*{u}\parallel\vb*{a}$ and $\vu*{v}$ rotated around $\vu*{u}$ by an angle $\theta$, see \cref{sec:chira}) as shown in \cref{fig:RITA_Ivt}. The observed intensities were normalized to the intensity at $\SI{1.75}{K}$, while the calculated intensities were normalized to the intensity at $\theta=\SI{90}{\degree}$. The matching of these normalized intensities indicates that the spin structure goes from a nearly pure cycloid state to a nearly pure helix state which is analogous to the zero field magnetic structure of iron based langasite. At $T=\SI{1.75}{K}$, the tilt angle of the spin rotation plane seems to increase linearly with the magnetic field, whereas the tilt starts at higher magnetic field for higher temperatures.

\begin{figure}[h]
    \begin{center}
    \includegraphics{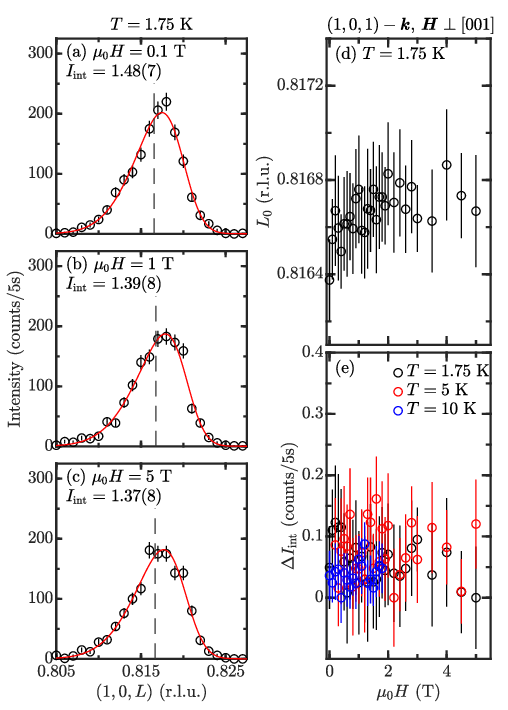}
    \end{center}
    \caption{The magnetic field is applied perpendicular to the $c$-axis. (a)-(c) Scans at $T=\SI{1.75}{K}$ along the $(00L)$ direction for different values of magnetic field. The mean position of the skewed Gaussian fits are shown in dashed lines and remain constant. (d) Summary of the field dependence of the propagation vector at $T=\SI{1.75}{K}$. (e) Integrated intensities of the magnetic peak $\Q=(1,0,1)-\vb*{k}$ as a function of the magnetic field. The zero-field intensity is subtracted from each respective curve for a clearer comparison of the data.}
    \label{fig:RITA_perp}
\end{figure}

The results are different when the magnetic field is rotated by 90° and applied in the $ab$-plane. In this case, the in-plane main axis of the cycloid will tend to be perpendicular to the magnetic field and the magnetic domains are simply reoriented in the $ab$-plane. As mentioned above, magnetic diffraction is not very sensitive to the direction of the in-plane main axis, because the intensities of the magnetic peaks do not change significantly between two directions of this axis. This is especially true for the magnetic peak $\Q=(1,0,1)-\vb*{k}$, where the measured intensities are constant as a function of the magnetic field (\cref{fig:RITA_perp}(a)). Contrary to previous thermodynamic magnetization measurements,\cite{werner201694} the in-plane reorientation of the spin structure cannot be detected in this experiment. The propagation vector also remains constant, within error, as a function of the magnetic field (\cref{fig:RITA_perp}(b)-(e)), indicating the absence of measurable magneto-elastic effects in this case.  We note that the difference in $L_{0}$ values between the two different field directions is an experimental artefact resulting from not being able to refine a zero offset in the scattering angle. This is due to only being able to measure a single Bragg peak owing to kinematic constraints imposed by the horizontal magnetic field geometry.

\section{Theory for an electric polarization}\label{sec:mf}

In their work, Kinoshita \textit{et al.} (Ref. \onlinecite{kinoshita2016117}) have measured the pyroelectric current in a single crystal of \mnsb\ along $\vu*{u}\parallel[1\bar{1}0]$ under a magnetic field rotating in the $(1\bar{1}0)$ plane. An electric polarization was measured for the magnetic field slightly off the $(ab)$-plane and was attributed to the selection of a tilted polar domain. This polarization is reversed when the magnetic field is applied on the other side of the $(ab)$-plane, favoring the opposite tilted polar domain. This mechanism relied on the tilted cycloid model considered as the ground state in \mnsb. In this section we discuss a phenomenological theory for the domain switching observed in Ref. \onlinecite{kinoshita2016117} under the application of a magnetic field in the absence of a zero-field tilt as discussed in our diffraction results outlined above.

As is the case for many compounds having a cycloidal magnetic structure, \mnsb\ is predicted to hold an electric polarization through the inverse DM interaction\cite{mostovoy200696} or spin-current induced\cite{katsura200595} mechanisms which predicts an electric polarization $\vb*{P}$ given by:

\begin{equation}
    \vb*{P}\propto\vb*{r}_{ij}\times(\vb*{S}_i\times\vb*{S}_j)
\end{equation}

\noindent which couples to the magnetic polarity $\vb*{P}\ut{m}$ in the phenomenological free energy through a term $\propto \lambda\vb*{P}\ut{m}\cdot\vb*{P}$. Therefore the electric polarization $\vb*{P}$ lies parallel or antiparallel to to the magnetic polarity $\vb*{P}\ut{m}$ depending on the sign of the coupling constant $\lambda$.\cite{johnson2013111} 

Using our definitions in \cref{eq:coschira} and \cref{eq:sinchira}, we can build trilinear invariants based upon Heisenberg exchange interactions from \cref{eq:invariant}:

\begin{equation}
\begin{aligned}
    \sigma\epsilon_H\epsilon_T &=& \sigma\etc\ett\sin^2(\theta) \\
    \sigma \vb*{P}\ut{m}\cdot\vb*{A}&=& \sigma\etc\ett\cos^2(\theta) 
    \end{aligned}
\end{equation}

\noindent Again, this shows the equivalence between the scalar and vector coupling schemes from iron langasite and \mnsb, and also the mixture of both with the spin rotation plane tilt angle $\theta$. These invariants imply that the polarization $\vb*{P}$, does not change sign with $\theta$, based on Heisenberg exchanges alone.

However, we can consider a uniform DM interaction with $\vb*{D}_{\alpha\beta}$ parallel to the $c$-axis. $\vb*{D}_{\alpha\beta}$ is a T-even axial vector that changes sign on exchange of indices. Its sign will also depend upon the structural chirality $\sigma$, hence we can write $\vb*{D}_{\alpha\beta} \propto \sigma \vb*{r}_{\alpha\beta}$ where $\vb*{r}_{\alpha\beta}$ is the bond vector between spins at sites $\alpha$ and $\beta$ along the $c$-axis. Following \cref{eq:sinchira}, the magnetic energy is then given by

\begin{equation}
   \begin{aligned}
    E\ut{DM} &=\vb*{D}_{\alpha\beta} \cdot (\vb*{S}_\alpha\times \vb*{S}_\beta)  \\
    &\propto\sigma\vb*{r}_{\alpha\beta}\cdot\vb*{V\ut{C}} \\ &\propto \sigma\eta\ut{C}\sin\theta
    \end{aligned} 
\end{equation}

\noindent Therefore, for a given structural domain with a fixed $\sigma$, when the sign of $\theta$ is inverted (through the application of a magnetic field) the uniform DM interaction will favor a change of sign of $\eta\ut{C}$ which in turn results in the sign of $\vb*{P}\ut{m}$ being inverted, from \cref{eq:coschira}. This will change the direction of the electric polarization $\vb*{P}$. The only condition for having a non-zero polarization is an imbalance of structural chiral domains for a given tilt angle $\theta$. This mechanism does not need the magnetic ground state to be tilted. Indeed, the ground state could be a pure cycloid stabilized by Heisenberg exchanges, where the anisotropy overcomes this small DM term. When an external magnetic field is applied slightly out of the ($ab$)-plane, this would overcome the anisotropy and tilt the spin rotation plane. In this case the DM term would lift the degeneracy of $\pm\etc$ domains, and give rise to a non-zero electric polarization for a given structural domain. We note that the DM term is allowed owing to the large distortion of the oxygen octahedra surrounding the Mn$^{2+}$ ions.\cite{Yosida:book}

From our diffraction data under magnetic field, we know that a small magnetic field (around 2 T) is sufficient to reorient the spin rotation plane perpendicular to the magnetic field, which is consistent with a small single-ion anisotropy in our compound and is consistent with the values used in Ref.~\onlinecite{kinoshita2016117} in their macroscopic measurement of electric polarization. However, Ref.~\onlinecite{kinoshita2016117} has considered the tilted cycloid ground state as essential for selecting the polar domains with an external magnetic field applied perpendicularly to the spin rotation plane. This explanation does not work in the case that \textit{magnetic} domains with polarity $\pm\etc$ have the exact same populations, because the overall polarization would compensate. As $\pm\etc$ domains are degenerate from Heisenberg model, our mechanism based on a uniform DM interaction is more general. In particular and in the context of MnSb$_{2}$O$_{6}$, this mechanism does not depend on a tilted ground state, and requires an imbalance in structural chiral domains and the underlying coupling between magnetic and structural chiralities. 

\section{Conclusion}

In this study, we have performed a combination of unpolarized and polarized neutron diffraction experiments on \mnsb. The study of the nuclear structure shows no evidence for the breaking of the paramagnetic crystallographic space group at the magnetic transition. The consideration of structural twins in our work shows that our single crystal is a non-racemic mixture of chiral structural domains. There is no evidence of a helicoidal magnetic ground state, but diffraction under magnetic field shows the possibility to manipulate the spin structure with low magnetic fields. Finally, we propose that a uniform DM interaction, combined with the underlying coupling between structural and magnetic chiralities, is sufficient to explain an electric polarization switching mechanism which was previously measured.

\begin{acknowledgements}

We would like to thank H. Lane and R. Ewings for helpful discussions.  We are grateful for funding from the Carnegie Trust for the Universities of Scotland, EPSRC, STFC, and the ILL.

\end{acknowledgements}


\begin{thebibliography}{53}%
\makeatletter
\providecommand \@ifxundefined [1]{%
 \@ifx{#1\undefined}
}%
\providecommand \@ifnum [1]{%
 \ifnum #1\expandafter \@firstoftwo
 \else \expandafter \@secondoftwo
 \fi
}%
\providecommand \@ifx [1]{%
 \ifx #1\expandafter \@firstoftwo
 \else \expandafter \@secondoftwo
 \fi
}%
\providecommand \natexlab [1]{#1}%
\providecommand \enquote  [1]{``#1''}%
\providecommand \bibnamefont  [1]{#1}%
\providecommand \bibfnamefont [1]{#1}%
\providecommand \citenamefont [1]{#1}%
\providecommand \href@noop [0]{\@secondoftwo}%
\providecommand \href [0]{\begingroup \@sanitize@url \@href}%
\providecommand \@href[1]{\@@startlink{#1}\@@href}%
\providecommand \@@href[1]{\endgroup#1\@@endlink}%
\providecommand \@sanitize@url [0]{\catcode `\\12\catcode `\$12\catcode
  `\&12\catcode `\#12\catcode `\^12\catcode `\_12\catcode `\%12\relax}%
\providecommand \@@startlink[1]{}%
\providecommand \@@endlink[0]{}%
\providecommand \url  [0]{\begingroup\@sanitize@url \@url }%
\providecommand \@url [1]{\endgroup\@href {#1}{\urlprefix }}%
\providecommand \urlprefix  [0]{URL }%
\providecommand \Eprint [0]{\href }%
\providecommand \doibase [0]{http://dx.doi.org/}%
\providecommand \selectlanguage [0]{\@gobble}%
\providecommand \bibinfo  [0]{\@secondoftwo}%
\providecommand \bibfield  [0]{\@secondoftwo}%
\providecommand \translation [1]{[#1]}%
\providecommand \BibitemOpen [0]{}%
\providecommand \bibitemStop [0]{}%
\providecommand \bibitemNoStop [0]{.\EOS\space}%
\providecommand \EOS [0]{\spacefactor3000\relax}%
\providecommand \BibitemShut  [1]{\csname bibitem#1\endcsname}%
\let\auto@bib@innerbib\@empty
\bibitem [{\citenamefont {Spaldin}\ and\ \citenamefont
  {Fiebig}(2005)}]{spaldin2005309}%
  \BibitemOpen
  \bibfield  {author} {\bibinfo {author} {\bibfnamefont {N.~A.}\ \bibnamefont
  {Spaldin}}\ and\ \bibinfo {author} {\bibfnamefont {M.}~\bibnamefont
  {Fiebig}},\ }\href {\doibase 10.1126/science.1113357} {\bibfield  {journal}
  {\bibinfo  {journal} {Science}\ }\textbf {\bibinfo {volume} {309}},\ \bibinfo
  {pages} {391} (\bibinfo {year} {2005})}\BibitemShut {NoStop}%
\bibitem [{\citenamefont {Cheong}\ and\ \citenamefont
  {Mostovoy}(2007)}]{cheong20076}%
  \BibitemOpen
  \bibfield  {author} {\bibinfo {author} {\bibfnamefont {S.-W.}\ \bibnamefont
  {Cheong}}\ and\ \bibinfo {author} {\bibfnamefont {M.}~\bibnamefont
  {Mostovoy}},\ }\href {\doibase 10.1038/nmat1804} {\bibfield  {journal}
  {\bibinfo  {journal} {Nat. Mater.}\ }\textbf {\bibinfo {volume} {6}},\
  \bibinfo {pages} {13} (\bibinfo {year} {2007})}\BibitemShut {NoStop}%
\bibitem [{\citenamefont {Tokura}\ \emph {et~al.}(2014)\citenamefont {Tokura},
  \citenamefont {Seki},\ and\ \citenamefont {Nagaosa}}]{tokura201477}%
  \BibitemOpen
  \bibfield  {author} {\bibinfo {author} {\bibfnamefont {Y.}~\bibnamefont
  {Tokura}}, \bibinfo {author} {\bibfnamefont {S.}~\bibnamefont {Seki}}, \ and\
  \bibinfo {author} {\bibfnamefont {N.}~\bibnamefont {Nagaosa}},\ }\href
  {\doibase 10.1088/0034-4885/77/7/076501} {\bibfield  {journal} {\bibinfo
  {journal} {Rep. Prog. Phys.}\ }\textbf {\bibinfo {volume} {77}},\ \bibinfo
  {pages} {076501} (\bibinfo {year} {2014})}\BibitemShut {NoStop}%
\bibitem [{\citenamefont {Dong}\ \emph {et~al.}(2015)\citenamefont {Dong},
  \citenamefont {Liu}, \citenamefont {Cheong},\ and\ \citenamefont
  {Ren}}]{dong201564}%
  \BibitemOpen
  \bibfield  {author} {\bibinfo {author} {\bibfnamefont {S.}~\bibnamefont
  {Dong}}, \bibinfo {author} {\bibfnamefont {J.-M.}\ \bibnamefont {Liu}},
  \bibinfo {author} {\bibfnamefont {S.-W.}\ \bibnamefont {Cheong}}, \ and\
  \bibinfo {author} {\bibfnamefont {Z.}~\bibnamefont {Ren}},\ }\href {\doibase
  10.1080/00018732.2015.1114338} {\bibfield  {journal} {\bibinfo  {journal}
  {Adv. Phys.}\ }\textbf {\bibinfo {volume} {64}},\ \bibinfo {pages} {519}
  (\bibinfo {year} {2015})}\BibitemShut {NoStop}%
\bibitem [{\citenamefont {Fiebig}\ \emph {et~al.}(2016)\citenamefont {Fiebig},
  \citenamefont {Lottermoser}, \citenamefont {Meier},\ and\ \citenamefont
  {Trassin}}]{fiebig20161a}%
  \BibitemOpen
  \bibfield  {author} {\bibinfo {author} {\bibfnamefont {M.}~\bibnamefont
  {Fiebig}}, \bibinfo {author} {\bibfnamefont {T.}~\bibnamefont {Lottermoser}},
  \bibinfo {author} {\bibfnamefont {D.}~\bibnamefont {Meier}}, \ and\ \bibinfo
  {author} {\bibfnamefont {M.}~\bibnamefont {Trassin}},\ }\href {\doibase
  10.1038/natrevmats.2016.46} {\bibfield  {journal} {\bibinfo  {journal} {Nat.
  Rev. Mater.}\ }\textbf {\bibinfo {volume} {1}},\ \bibinfo {pages} {16046}
  (\bibinfo {year} {2016})}\BibitemShut {NoStop}%
\bibitem [{\citenamefont {Kimura}\ \emph {et~al.}(2003)\citenamefont {Kimura},
  \citenamefont {Goto}, \citenamefont {Shintani}, \citenamefont {Ishizaka},
  \citenamefont {Arima},\ and\ \citenamefont {Tokura}}]{kimura2003426}%
  \BibitemOpen
  \bibfield  {author} {\bibinfo {author} {\bibfnamefont {T.}~\bibnamefont
  {Kimura}}, \bibinfo {author} {\bibfnamefont {T.}~\bibnamefont {Goto}},
  \bibinfo {author} {\bibfnamefont {H.}~\bibnamefont {Shintani}}, \bibinfo
  {author} {\bibfnamefont {K.}~\bibnamefont {Ishizaka}}, \bibinfo {author}
  {\bibfnamefont {T.}~\bibnamefont {Arima}}, \ and\ \bibinfo {author}
  {\bibfnamefont {Y.}~\bibnamefont {Tokura}},\ }\href {\doibase
  10.1038/nature02018} {\bibfield  {journal} {\bibinfo  {journal} {Nature}\
  }\textbf {\bibinfo {volume} {426}},\ \bibinfo {pages} {55} (\bibinfo {year}
  {2003})}\BibitemShut {NoStop}%
\bibitem [{\citenamefont {Kenzelmann}\ \emph {et~al.}(2005)\citenamefont
  {Kenzelmann}, \citenamefont {Harris}, \citenamefont {Jonas}, \citenamefont
  {Broholm}, \citenamefont {Schefer}, \citenamefont {Kim}, \citenamefont
  {Zhang}, \citenamefont {Cheong}, \citenamefont {Vajk},\ and\ \citenamefont
  {Lynn}}]{kenzelmann200595}%
  \BibitemOpen
  \bibfield  {author} {\bibinfo {author} {\bibfnamefont {M.}~\bibnamefont
  {Kenzelmann}}, \bibinfo {author} {\bibfnamefont {A.~B.}\ \bibnamefont
  {Harris}}, \bibinfo {author} {\bibfnamefont {S.}~\bibnamefont {Jonas}},
  \bibinfo {author} {\bibfnamefont {C.}~\bibnamefont {Broholm}}, \bibinfo
  {author} {\bibfnamefont {J.}~\bibnamefont {Schefer}}, \bibinfo {author}
  {\bibfnamefont {S.~B.}\ \bibnamefont {Kim}}, \bibinfo {author} {\bibfnamefont
  {C.~L.}\ \bibnamefont {Zhang}}, \bibinfo {author} {\bibfnamefont {S.-W.}\
  \bibnamefont {Cheong}}, \bibinfo {author} {\bibfnamefont {O.~P.}\
  \bibnamefont {Vajk}}, \ and\ \bibinfo {author} {\bibfnamefont {J.~W.}\
  \bibnamefont {Lynn}},\ }\href {\doibase 10.1103/PhysRevLett.95.087206}
  {\bibfield  {journal} {\bibinfo  {journal} {Phys. Rev. Lett.}\ }\textbf
  {\bibinfo {volume} {95}},\ \bibinfo {pages} {087206} (\bibinfo {year}
  {2005})}\BibitemShut {NoStop}%
\bibitem [{\citenamefont {Katsura}\ \emph {et~al.}(2005)\citenamefont
  {Katsura}, \citenamefont {Nagaosa},\ and\ \citenamefont
  {Balatsky}}]{katsura200595}%
  \BibitemOpen
  \bibfield  {author} {\bibinfo {author} {\bibfnamefont {H.}~\bibnamefont
  {Katsura}}, \bibinfo {author} {\bibfnamefont {N.}~\bibnamefont {Nagaosa}}, \
  and\ \bibinfo {author} {\bibfnamefont {A.~V.}\ \bibnamefont {Balatsky}},\
  }\href {\doibase 10.1103/PhysRevLett.95.057205} {\bibfield  {journal}
  {\bibinfo  {journal} {Phys. Rev. Lett.}\ }\textbf {\bibinfo {volume} {95}},\
  \bibinfo {pages} {057205} (\bibinfo {year} {2005})}\BibitemShut {NoStop}%
\bibitem [{\citenamefont {Mostovoy}(2006)}]{mostovoy200696}%
  \BibitemOpen
  \bibfield  {author} {\bibinfo {author} {\bibfnamefont {M.}~\bibnamefont
  {Mostovoy}},\ }\href {\doibase 10.1103/PhysRevLett.96.067601} {\bibfield
  {journal} {\bibinfo  {journal} {Phys. Rev. Lett.}\ }\textbf {\bibinfo
  {volume} {96}},\ \bibinfo {pages} {067601} (\bibinfo {year}
  {2006})}\BibitemShut {NoStop}%
\bibitem [{\citenamefont {Kimura}(2007)}]{kimura200737}%
  \BibitemOpen
  \bibfield  {author} {\bibinfo {author} {\bibfnamefont {T.}~\bibnamefont
  {Kimura}},\ }\href {\doibase 10.1146/annurev.matsci.37.052506.084259}
  {\bibfield  {journal} {\bibinfo  {journal} {Annu. Rev. Mater. Res.}\ }\textbf
  {\bibinfo {volume} {37}},\ \bibinfo {pages} {387} (\bibinfo {year}
  {2007})}\BibitemShut {NoStop}%
\bibitem [{\citenamefont {Marty}\ \emph {et~al.}(2008)\citenamefont {Marty},
  \citenamefont {Simonet}, \citenamefont {Ressouche}, \citenamefont {Ballou},
  \citenamefont {Lejay},\ and\ \citenamefont {Bordet}}]{marty2008101}%
  \BibitemOpen
  \bibfield  {author} {\bibinfo {author} {\bibfnamefont {K.}~\bibnamefont
  {Marty}}, \bibinfo {author} {\bibfnamefont {V.}~\bibnamefont {Simonet}},
  \bibinfo {author} {\bibfnamefont {E.}~\bibnamefont {Ressouche}}, \bibinfo
  {author} {\bibfnamefont {R.}~\bibnamefont {Ballou}}, \bibinfo {author}
  {\bibfnamefont {P.}~\bibnamefont {Lejay}}, \ and\ \bibinfo {author}
  {\bibfnamefont {P.}~\bibnamefont {Bordet}},\ }\href {\doibase
  10.1103/PhysRevLett.101.247201} {\bibfield  {journal} {\bibinfo  {journal}
  {Phys. Rev. Lett.}\ }\textbf {\bibinfo {volume} {101}},\ \bibinfo {pages}
  {247201} (\bibinfo {year} {2008})}\BibitemShut {NoStop}%
\bibitem [{\citenamefont {Loire}\ \emph {et~al.}(2011)\citenamefont {Loire},
  \citenamefont {Simonet}, \citenamefont {Petit}, \citenamefont {Marty},
  \citenamefont {Bordet}, \citenamefont {Lejay}, \citenamefont {Ollivier},
  \citenamefont {Enderle}, \citenamefont {Steffens}, \citenamefont {Ressouche},
  \citenamefont {Zorko},\ and\ \citenamefont {Ballou}}]{loire2011106}%
  \BibitemOpen
  \bibfield  {author} {\bibinfo {author} {\bibfnamefont {M.}~\bibnamefont
  {Loire}}, \bibinfo {author} {\bibfnamefont {V.}~\bibnamefont {Simonet}},
  \bibinfo {author} {\bibfnamefont {S.}~\bibnamefont {Petit}}, \bibinfo
  {author} {\bibfnamefont {K.}~\bibnamefont {Marty}}, \bibinfo {author}
  {\bibfnamefont {P.}~\bibnamefont {Bordet}}, \bibinfo {author} {\bibfnamefont
  {P.}~\bibnamefont {Lejay}}, \bibinfo {author} {\bibfnamefont
  {J.}~\bibnamefont {Ollivier}}, \bibinfo {author} {\bibfnamefont
  {M.}~\bibnamefont {Enderle}}, \bibinfo {author} {\bibfnamefont
  {P.}~\bibnamefont {Steffens}}, \bibinfo {author} {\bibfnamefont
  {E.}~\bibnamefont {Ressouche}}, \bibinfo {author} {\bibfnamefont
  {A.}~\bibnamefont {Zorko}}, \ and\ \bibinfo {author} {\bibfnamefont
  {R.}~\bibnamefont {Ballou}},\ }\href {\doibase
  10.1103/PhysRevLett.106.207201} {\bibfield  {journal} {\bibinfo  {journal}
  {Phys. Rev. Lett.}\ }\textbf {\bibinfo {volume} {106}},\ \bibinfo {pages}
  {207201} (\bibinfo {year} {2011})}\BibitemShut {NoStop}%
\bibitem [{\citenamefont {Stock}\ \emph {et~al.}(2011)\citenamefont {Stock},
  \citenamefont {Chapon}, \citenamefont {Schneidewind}, \citenamefont {Su},
  \citenamefont {Radaelli}, \citenamefont {McMorrow}, \citenamefont {Bombardi},
  \citenamefont {Lee},\ and\ \citenamefont {Cheong}}]{stock201183}%
  \BibitemOpen
  \bibfield  {author} {\bibinfo {author} {\bibfnamefont {C.}~\bibnamefont
  {Stock}}, \bibinfo {author} {\bibfnamefont {L.~C.}\ \bibnamefont {Chapon}},
  \bibinfo {author} {\bibfnamefont {A.}~\bibnamefont {Schneidewind}}, \bibinfo
  {author} {\bibfnamefont {Y.}~\bibnamefont {Su}}, \bibinfo {author}
  {\bibfnamefont {P.~G.}\ \bibnamefont {Radaelli}}, \bibinfo {author}
  {\bibfnamefont {D.~F.}\ \bibnamefont {McMorrow}}, \bibinfo {author}
  {\bibfnamefont {A.}~\bibnamefont {Bombardi}}, \bibinfo {author}
  {\bibfnamefont {N.}~\bibnamefont {Lee}}, \ and\ \bibinfo {author}
  {\bibfnamefont {S.-W.}\ \bibnamefont {Cheong}},\ }\href {\doibase
  10.1103/PhysRevB.83.104426} {\bibfield  {journal} {\bibinfo  {journal} {Phys.
  Rev. B}\ }\textbf {\bibinfo {volume} {83}},\ \bibinfo {pages} {104426}
  (\bibinfo {year} {2011})}\BibitemShut {NoStop}%
\bibitem [{\citenamefont {Chaix}\ \emph {et~al.}(2016)\citenamefont {Chaix},
  \citenamefont {Ballou}, \citenamefont {Cano}, \citenamefont {Petit},
  \citenamefont {{de Brion}}, \citenamefont {Ollivier}, \citenamefont
  {Regnault}, \citenamefont {Ressouche}, \citenamefont {Constable},
  \citenamefont {Colin}, \citenamefont {Zorko}, \citenamefont {Scagnoli},
  \citenamefont {Balay}, \citenamefont {Lejay},\ and\ \citenamefont
  {Simonet}}]{chaix2016}%
  \BibitemOpen
  \bibfield  {author} {\bibinfo {author} {\bibfnamefont {L.}~\bibnamefont
  {Chaix}}, \bibinfo {author} {\bibfnamefont {R.}~\bibnamefont {Ballou}},
  \bibinfo {author} {\bibfnamefont {A.}~\bibnamefont {Cano}}, \bibinfo {author}
  {\bibfnamefont {S.}~\bibnamefont {Petit}}, \bibinfo {author} {\bibfnamefont
  {S.}~\bibnamefont {{de Brion}}}, \bibinfo {author} {\bibfnamefont
  {J.}~\bibnamefont {Ollivier}}, \bibinfo {author} {\bibfnamefont {L.-P.}\
  \bibnamefont {Regnault}}, \bibinfo {author} {\bibfnamefont {E.}~\bibnamefont
  {Ressouche}}, \bibinfo {author} {\bibfnamefont {E.}~\bibnamefont
  {Constable}}, \bibinfo {author} {\bibfnamefont {C.~V.}\ \bibnamefont
  {Colin}}, \bibinfo {author} {\bibfnamefont {A.}~\bibnamefont {Zorko}},
  \bibinfo {author} {\bibfnamefont {V.}~\bibnamefont {Scagnoli}}, \bibinfo
  {author} {\bibfnamefont {J.}~\bibnamefont {Balay}}, \bibinfo {author}
  {\bibfnamefont {P.}~\bibnamefont {Lejay}}, \ and\ \bibinfo {author}
  {\bibfnamefont {V.}~\bibnamefont {Simonet}},\ }\href {\doibase
  10.1103/PhysRevB.93.214419} {\bibfield  {journal} {\bibinfo  {journal} {Phys.
  Rev. B}\ }\textbf {\bibinfo {volume} {93}},\ \bibinfo {pages} {214419}
  (\bibinfo {year} {2016})}\BibitemShut {NoStop}%
\bibitem [{\citenamefont {Stock}\ \emph {et~al.}(2019)\citenamefont {Stock},
  \citenamefont {Johnson}, \citenamefont {{Giles-Donovan}}, \citenamefont
  {Songvilay}, \citenamefont {{Rodriguez-Rivera}}, \citenamefont {Lee},
  \citenamefont {Xu}, \citenamefont {Radaelli}, \citenamefont {Chapon},
  \citenamefont {Bombardi}, \citenamefont {Cochran}, \citenamefont
  {Niedermayer}, \citenamefont {Schneidewind}, \citenamefont {Husges},
  \citenamefont {Lu}, \citenamefont {Meng},\ and\ \citenamefont
  {Cheong}}]{stock2019100}%
  \BibitemOpen
  \bibfield  {author} {\bibinfo {author} {\bibfnamefont {C.}~\bibnamefont
  {Stock}}, \bibinfo {author} {\bibfnamefont {R.~D.}\ \bibnamefont {Johnson}},
  \bibinfo {author} {\bibfnamefont {N.}~\bibnamefont {{Giles-Donovan}}},
  \bibinfo {author} {\bibfnamefont {M.}~\bibnamefont {Songvilay}}, \bibinfo
  {author} {\bibfnamefont {J.~A.}\ \bibnamefont {{Rodriguez-Rivera}}}, \bibinfo
  {author} {\bibfnamefont {N.}~\bibnamefont {Lee}}, \bibinfo {author}
  {\bibfnamefont {X.}~\bibnamefont {Xu}}, \bibinfo {author} {\bibfnamefont
  {P.~G.}\ \bibnamefont {Radaelli}}, \bibinfo {author} {\bibfnamefont {L.~C.}\
  \bibnamefont {Chapon}}, \bibinfo {author} {\bibfnamefont {A.}~\bibnamefont
  {Bombardi}}, \bibinfo {author} {\bibfnamefont {S.}~\bibnamefont {Cochran}},
  \bibinfo {author} {\bibfnamefont {C.}~\bibnamefont {Niedermayer}}, \bibinfo
  {author} {\bibfnamefont {A.}~\bibnamefont {Schneidewind}}, \bibinfo {author}
  {\bibfnamefont {Z.}~\bibnamefont {Husges}}, \bibinfo {author} {\bibfnamefont
  {Z.}~\bibnamefont {Lu}}, \bibinfo {author} {\bibfnamefont {S.}~\bibnamefont
  {Meng}}, \ and\ \bibinfo {author} {\bibfnamefont {S.-W.}\ \bibnamefont
  {Cheong}},\ }\href {\doibase 10.1103/PhysRevB.100.134429} {\bibfield
  {journal} {\bibinfo  {journal} {Phys. Rev. B}\ }\textbf {\bibinfo {volume}
  {100}},\ \bibinfo {pages} {134429} (\bibinfo {year} {2019})}\BibitemShut
  {NoStop}%
\bibitem [{\citenamefont {Ramakrishnan}\ \emph {et~al.}(2019)\citenamefont
  {Ramakrishnan}, \citenamefont {Constable}, \citenamefont {Cano},
  \citenamefont {Mostovoy}, \citenamefont {White}, \citenamefont {Gurung},
  \citenamefont {Schierle}, \citenamefont {de~Brion}, \citenamefont {Colin},
  \citenamefont {Gay}, \citenamefont {Lejay}, \citenamefont {Ressouche},
  \citenamefont {Weschke}, \citenamefont {Scagnoli}, \citenamefont {Ballou},
  \citenamefont {Simonet},\ and\ \citenamefont {Staub}}]{ramakrishnan20194}%
  \BibitemOpen
  \bibfield  {author} {\bibinfo {author} {\bibfnamefont {M.}~\bibnamefont
  {Ramakrishnan}}, \bibinfo {author} {\bibfnamefont {E.}~\bibnamefont
  {Constable}}, \bibinfo {author} {\bibfnamefont {A.}~\bibnamefont {Cano}},
  \bibinfo {author} {\bibfnamefont {M.}~\bibnamefont {Mostovoy}}, \bibinfo
  {author} {\bibfnamefont {J.~S.}\ \bibnamefont {White}}, \bibinfo {author}
  {\bibfnamefont {N.}~\bibnamefont {Gurung}}, \bibinfo {author} {\bibfnamefont
  {E.}~\bibnamefont {Schierle}}, \bibinfo {author} {\bibfnamefont
  {S.}~\bibnamefont {de~Brion}}, \bibinfo {author} {\bibfnamefont {C.~V.}\
  \bibnamefont {Colin}}, \bibinfo {author} {\bibfnamefont {F.}~\bibnamefont
  {Gay}}, \bibinfo {author} {\bibfnamefont {P.}~\bibnamefont {Lejay}}, \bibinfo
  {author} {\bibfnamefont {E.}~\bibnamefont {Ressouche}}, \bibinfo {author}
  {\bibfnamefont {E.}~\bibnamefont {Weschke}}, \bibinfo {author} {\bibfnamefont
  {V.}~\bibnamefont {Scagnoli}}, \bibinfo {author} {\bibfnamefont
  {R.}~\bibnamefont {Ballou}}, \bibinfo {author} {\bibfnamefont
  {V.}~\bibnamefont {Simonet}}, \ and\ \bibinfo {author} {\bibfnamefont
  {U.}~\bibnamefont {Staub}},\ }\href {\doibase 10.1038/s41535-019-0199-3}
  {\bibfield  {journal} {\bibinfo  {journal} {npj Quantum Mater.}\ }\textbf
  {\bibinfo {volume} {4}},\ \bibinfo {pages} {60} (\bibinfo {year}
  {2019})}\BibitemShut {NoStop}%
\bibitem [{\citenamefont {Johnson}\ \emph {et~al.}(2013)\citenamefont
  {Johnson}, \citenamefont {Cao}, \citenamefont {Chapon}, \citenamefont
  {Fabrizi}, \citenamefont {Perks}, \citenamefont {Manuel}, \citenamefont
  {Yang}, \citenamefont {Oh}, \citenamefont {Cheong},\ and\ \citenamefont
  {Radaelli}}]{johnson2013111}%
  \BibitemOpen
  \bibfield  {author} {\bibinfo {author} {\bibfnamefont {R.~D.}\ \bibnamefont
  {Johnson}}, \bibinfo {author} {\bibfnamefont {K.}~\bibnamefont {Cao}},
  \bibinfo {author} {\bibfnamefont {L.~C.}\ \bibnamefont {Chapon}}, \bibinfo
  {author} {\bibfnamefont {F.}~\bibnamefont {Fabrizi}}, \bibinfo {author}
  {\bibfnamefont {N.}~\bibnamefont {Perks}}, \bibinfo {author} {\bibfnamefont
  {P.}~\bibnamefont {Manuel}}, \bibinfo {author} {\bibfnamefont {J.~J.}\
  \bibnamefont {Yang}}, \bibinfo {author} {\bibfnamefont {Y.~S.}\ \bibnamefont
  {Oh}}, \bibinfo {author} {\bibfnamefont {S.-W.}\ \bibnamefont {Cheong}}, \
  and\ \bibinfo {author} {\bibfnamefont {P.~G.}\ \bibnamefont {Radaelli}},\
  }\href {\doibase 10.1103/PhysRevLett.111.017202} {\bibfield  {journal}
  {\bibinfo  {journal} {Phys. Rev. Lett.}\ }\textbf {\bibinfo {volume} {111}},\
  \bibinfo {pages} {017202} (\bibinfo {year} {2013})}\BibitemShut {NoStop}%
\bibitem [{\citenamefont {Reimers}\ and\ \citenamefont
  {Greedan}(1989)}]{reimers1989}%
  \BibitemOpen
  \bibfield  {author} {\bibinfo {author} {\bibfnamefont {J.~N.}\ \bibnamefont
  {Reimers}}\ and\ \bibinfo {author} {\bibfnamefont {J.~E.}\ \bibnamefont
  {Greedan}},\ }\href {\doibase 10.1016/0022-4596(89)90273-9} {\bibfield
  {journal} {\bibinfo  {journal} {J. Solid State Chem.}\ }\textbf {\bibinfo
  {volume} {79}},\ \bibinfo {pages} {263} (\bibinfo {year} {1989})}\BibitemShut
  {NoStop}%
\bibitem [{\citenamefont {Werner}\ \emph {et~al.}(2016)\citenamefont {Werner},
  \citenamefont {Koo}, \citenamefont {Klingeler}, \citenamefont {Vasiliev},
  \citenamefont {Ovchenkov}, \citenamefont {Polovkova}, \citenamefont
  {Raganyan},\ and\ \citenamefont {Zvereva}}]{werner201694}%
  \BibitemOpen
  \bibfield  {author} {\bibinfo {author} {\bibfnamefont {J.}~\bibnamefont
  {Werner}}, \bibinfo {author} {\bibfnamefont {C.}~\bibnamefont {Koo}},
  \bibinfo {author} {\bibfnamefont {R.}~\bibnamefont {Klingeler}}, \bibinfo
  {author} {\bibfnamefont {A.~N.}\ \bibnamefont {Vasiliev}}, \bibinfo {author}
  {\bibfnamefont {Y.~A.}\ \bibnamefont {Ovchenkov}}, \bibinfo {author}
  {\bibfnamefont {A.~S.}\ \bibnamefont {Polovkova}}, \bibinfo {author}
  {\bibfnamefont {G.~V.}\ \bibnamefont {Raganyan}}, \ and\ \bibinfo {author}
  {\bibfnamefont {E.~A.}\ \bibnamefont {Zvereva}},\ }\href {\doibase
  10.1103/PhysRevB.94.104408} {\bibfield  {journal} {\bibinfo  {journal} {Phys.
  Rev. B}\ }\textbf {\bibinfo {volume} {94}},\ \bibinfo {pages} {104408}
  (\bibinfo {year} {2016})}\BibitemShut {NoStop}%
\bibitem [{\citenamefont {Aizu}(1970)}]{aizu19702}%
  \BibitemOpen
  \bibfield  {author} {\bibinfo {author} {\bibfnamefont {K.}~\bibnamefont
  {Aizu}},\ }\href {\doibase 10.1103/PhysRevB.2.754} {\bibfield  {journal}
  {\bibinfo  {journal} {Phys. Rev. B}\ }\textbf {\bibinfo {volume} {2}},\
  \bibinfo {pages} {754} (\bibinfo {year} {1970})}\BibitemShut {NoStop}%
\bibitem [{\citenamefont {Kinoshita}\ \emph {et~al.}(2016)\citenamefont
  {Kinoshita}, \citenamefont {Seki}, \citenamefont {Sato}, \citenamefont
  {Nambu}, \citenamefont {Hong}, \citenamefont {Matsuda}, \citenamefont {Cao},
  \citenamefont {Ishiwata},\ and\ \citenamefont {Tokura}}]{kinoshita2016117}%
  \BibitemOpen
  \bibfield  {author} {\bibinfo {author} {\bibfnamefont {M.}~\bibnamefont
  {Kinoshita}}, \bibinfo {author} {\bibfnamefont {S.}~\bibnamefont {Seki}},
  \bibinfo {author} {\bibfnamefont {T.~J.}\ \bibnamefont {Sato}}, \bibinfo
  {author} {\bibfnamefont {Y.}~\bibnamefont {Nambu}}, \bibinfo {author}
  {\bibfnamefont {T.}~\bibnamefont {Hong}}, \bibinfo {author} {\bibfnamefont
  {M.}~\bibnamefont {Matsuda}}, \bibinfo {author} {\bibfnamefont {H.~B.}\
  \bibnamefont {Cao}}, \bibinfo {author} {\bibfnamefont {S.}~\bibnamefont
  {Ishiwata}}, \ and\ \bibinfo {author} {\bibfnamefont {Y.}~\bibnamefont
  {Tokura}},\ }\href {\doibase 10.1103/PhysRevLett.117.047201} {\bibfield
  {journal} {\bibinfo  {journal} {Phys. Rev. Lett.}\ }\textbf {\bibinfo
  {volume} {117}},\ \bibinfo {pages} {047201} (\bibinfo {year}
  {2016})}\BibitemShut {NoStop}%
\bibitem [{\citenamefont {Momma}\ and\ \citenamefont
  {Izumi}(2011)}]{momma201144}%
  \BibitemOpen
  \bibfield  {author} {\bibinfo {author} {\bibfnamefont {K.}~\bibnamefont
  {Momma}}\ and\ \bibinfo {author} {\bibfnamefont {F.}~\bibnamefont {Izumi}},\
  }\href {\doibase 10.1107/S0021889811038970} {\bibfield  {journal} {\bibinfo
  {journal} {J. Appl. Crystallogr.}\ }\textbf {\bibinfo {volume} {44}},\
  \bibinfo {pages} {1272} (\bibinfo {year} {2011})}\BibitemShut {NoStop}%
\bibitem [{\citenamefont {Qureshi}(2019)}]{qureshi201952}%
  \BibitemOpen
  \bibfield  {author} {\bibinfo {author} {\bibfnamefont {N.}~\bibnamefont
  {Qureshi}},\ }\href {\doibase 10.1107/S1600576718016084} {\bibfield
  {journal} {\bibinfo  {journal} {J. Appl. Crystallogr.}\ }\textbf {\bibinfo
  {volume} {52}},\ \bibinfo {pages} {175} (\bibinfo {year} {2019})}\BibitemShut
  {NoStop}%
\bibitem [{\citenamefont {Nakua}\ and\ \citenamefont
  {Greedan}(1995)}]{nakua1995154}%
  \BibitemOpen
  \bibfield  {author} {\bibinfo {author} {\bibfnamefont {A.~M.}\ \bibnamefont
  {Nakua}}\ and\ \bibinfo {author} {\bibfnamefont {J.~E.}\ \bibnamefont
  {Greedan}},\ }\href {\doibase 10.1016/0022-0248(95)00217-0} {\bibfield
  {journal} {\bibinfo  {journal} {J. Cryst. Growth}\ }\textbf {\bibinfo
  {volume} {154}},\ \bibinfo {pages} {334} (\bibinfo {year}
  {1995})}\BibitemShut {NoStop}%
\bibitem []{MSO2021D9}%
  \BibitemOpen
  \bibfield  {author} {\bibinfo {author} {\bibfnamefont {C.}~\bibnamefont
  {Stock}}, \bibinfo {author} {\bibfnamefont {O.~R.}\ \bibnamefont
  {Fabelo~Rosa}}, \bibinfo {author} {\bibfnamefont {N.}~\bibnamefont
  {Qureshi}}, \ and\ \bibinfo {author} {\bibfnamefont {M.}~\bibnamefont
  {Songvilay}},\ }
  \bibinfo {note} {Institut Laue-Langevin (ILL) }
 (\bibinfo {year} {2021}),\
  \bibinfo {note} {doi:}
  \href {\doibase 10.5291/ILL-DATA.5-11-439}
  {\bibinfo {note} {10.5291/ILL-DATA.5-11-439}}\  \BibitemShut {NoStop}%
\bibitem [{\citenamefont {Chan}\ \emph
  {et~al.}(2021{\natexlab{a}})\citenamefont {Chan}, \citenamefont {Beauvois},
  \citenamefont {Qureshi}, \citenamefont {Rodriguez~Velamazan}, \citenamefont
  {Songvilay}, \citenamefont {Stock},\ and\ \citenamefont
  {Stunault}}]{MSO2021D10}%
  \BibitemOpen
  \bibfield  {author} {\bibinfo {author} {\bibfnamefont {E.}~\bibnamefont
  {Chan}}, \bibinfo {author} {\bibfnamefont {K.}~\bibnamefont {Beauvois}},
  \bibinfo {author} {\bibfnamefont {N.}~\bibnamefont {Qureshi}}, \bibinfo
  {author} {\bibfnamefont {J.~A.}\ \bibnamefont {Rodriguez~Velamazan}},
  \bibinfo {author} {\bibfnamefont {M.}~\bibnamefont {Songvilay}}, \bibinfo
  {author} {\bibfnamefont {C.}~\bibnamefont {Stock}}, \ and\ \bibinfo {author}
  {\bibfnamefont {A.}~\bibnamefont {Stunault}},\ }
    \bibinfo {note} {Institut Laue-Langevin (ILL) }
 (\bibinfo {year} {2021}),\
  \bibinfo {note} {doi:}
  \href {\doibase
  10.5291/ILL-DATA.5-41-1169} { {\bibinfo {title} {10.5291/ILL-DATA.5-41-1169}}}\BibitemShut
  {NoStop}%
\bibitem [{\citenamefont {Tasset}\ \emph {et~al.}(1999)\citenamefont {Tasset},
  \citenamefont {Brown}, \citenamefont {{Leli{\`e}vre-Berna}}, \citenamefont
  {Roberts}, \citenamefont {Pujol}, \citenamefont {Allibon},\ and\
  \citenamefont {{Bourgeat-Lami}}}]{tasset1999267-268a}%
  \BibitemOpen
  \bibfield  {author} {\bibinfo {author} {\bibfnamefont {F.}~\bibnamefont
  {Tasset}}, \bibinfo {author} {\bibfnamefont {P.~J.}~\bibnamefont {Brown}},
  \bibinfo {author} {\bibfnamefont {E.}~\bibnamefont {{Leli{\`e}vre-Berna}}},
  \bibinfo {author} {\bibfnamefont {T.}~\bibnamefont {Roberts}}, \bibinfo
  {author} {\bibfnamefont {S.}~\bibnamefont {Pujol}}, \bibinfo {author}
  {\bibfnamefont {J.}~\bibnamefont {Allibon}}, \ and\ \bibinfo {author}
  {\bibfnamefont {E.}~\bibnamefont {{Bourgeat-Lami}}},\ }\href {\doibase
  10.1016/S0921-4526(99)00029-0} {\bibfield  {journal} {\bibinfo  {journal}
  {Phys. B}\ }\textbf {\bibinfo {volume} {267--268}},\ \bibinfo {pages} {69}
  (\bibinfo {year} {1999})}\BibitemShut {NoStop}%
\bibitem [{\citenamefont {Stock}\ \emph {et~al.}(2017)\citenamefont {Stock},
  \citenamefont {Chapon}, \citenamefont {Pasztorova}, \citenamefont {Qureshi},
  \citenamefont {Songvilay},\ and\ \citenamefont {Stunault}}]{MSO2017D3}%
  \BibitemOpen
  \bibfield  {author} {\bibinfo {author} {\bibfnamefont {C.}~\bibnamefont
  {Stock}}, \bibinfo {author} {\bibfnamefont {L.}~\bibnamefont {Chapon}},
  \bibinfo {author} {\bibfnamefont {J.}~\bibnamefont {Pasztorova}}, \bibinfo
  {author} {\bibfnamefont {N.}~\bibnamefont {Qureshi}}, \bibinfo {author}
  {\bibfnamefont {M.}~\bibnamefont {Songvilay}}, \ and\ \bibinfo {author}
  {\bibfnamefont {A.}~\bibnamefont {Stunault}},\ }
    \bibinfo {note} {Institut Laue-Langevin (ILL) }
 (\bibinfo {year} {2017}),\
  \bibinfo {note} {doi:}
  \href {\doibase
  10.5291/ILL-DATA.5-54-237} {10.5291/ILL-DATA.5-54-237}
  \BibitemShut {NoStop}%
\bibitem [{\citenamefont {Chan}\ \emph
  {et~al.}(2021{\natexlab{b}})\citenamefont {Chan}, \citenamefont {Qureshi},
  \citenamefont {Ritter},\ and\ \citenamefont {Stock}}]{MSO2021D20}%
  \BibitemOpen
  \bibfield  {author} {\bibinfo {author} {\bibfnamefont {E.}~\bibnamefont
  {Chan}}, \bibinfo {author} {\bibfnamefont {N.}~\bibnamefont {Qureshi}},
  \bibinfo {author} {\bibfnamefont {C.}~\bibnamefont {Ritter}}, \ and\ \bibinfo
  {author} {\bibfnamefont {C.}~\bibnamefont {Stock}},\ }
  \bibinfo {note} {Institut Laue-Langevin (ILL) }
 (\bibinfo {year} {2021}),\
  \bibinfo {note} {doi:}
  \href {\doibase
  10.5291/ILL-DATA.5-31-2794} {10.5291/ILL-DATA.5-31-2794}\BibitemShut
  {NoStop}%
\bibitem [{\citenamefont {Schwinger}(1948)}]{schwinger194873}%
  \BibitemOpen
  \bibfield  {author} {\bibinfo {author} {\bibfnamefont {J.}~\bibnamefont
  {Schwinger}},\ }\href {\doibase 10.1103/PhysRev.73.407} {\bibfield  {journal}
  {\bibinfo  {journal} {Phys. Rev.}\ }\textbf {\bibinfo {volume} {73}},\
  \bibinfo {pages} {407} (\bibinfo {year} {1948})}\BibitemShut {NoStop}%
\bibitem [{\citenamefont {Felcher}\ and\ \citenamefont
  {Peterson}(1975)}]{felcher197531}%
  \BibitemOpen
  \bibfield  {author} {\bibinfo {author} {\bibfnamefont {G.~P.}\ \bibnamefont
  {Felcher}}\ and\ \bibinfo {author} {\bibfnamefont {S.~W.}\ \bibnamefont
  {Peterson}},\ }\href {\doibase 10.1107/S0567739475000149} {\bibfield
  {journal} {\bibinfo  {journal} {Acta Crystallogr. A}\ }\textbf {\bibinfo
  {volume} {31}},\ \bibinfo {pages} {76} (\bibinfo {year} {1975})}\BibitemShut
  {NoStop}%
\bibitem [{\citenamefont {Qureshi}\ \emph {et~al.}(2020)\citenamefont
  {Qureshi}, \citenamefont {Bombardi}, \citenamefont {Picozzi}, \citenamefont
  {Barone}, \citenamefont {{Leli{\`e}vre-Berna}}, \citenamefont {Xu},
  \citenamefont {Stock}, \citenamefont {McMorrow}, \citenamefont {Hearmon},
  \citenamefont {Fabrizi}, \citenamefont {Radaelli}, \citenamefont {Cheong},\
  and\ \citenamefont {Chapon}}]{qureshi2020102}%
  \BibitemOpen
  \bibfield  {author} {\bibinfo {author} {\bibfnamefont {N.}~\bibnamefont
  {Qureshi}}, \bibinfo {author} {\bibfnamefont {A.}~\bibnamefont {Bombardi}},
  \bibinfo {author} {\bibfnamefont {S.}~\bibnamefont {Picozzi}}, \bibinfo
  {author} {\bibfnamefont {P.}~\bibnamefont {Barone}}, \bibinfo {author}
  {\bibfnamefont {E.}~\bibnamefont {{Leli{\`e}vre-Berna}}}, \bibinfo {author}
  {\bibfnamefont {X.}~\bibnamefont {Xu}}, \bibinfo {author} {\bibfnamefont
  {C.}~\bibnamefont {Stock}}, \bibinfo {author} {\bibfnamefont {D.~F.}\
  \bibnamefont {McMorrow}}, \bibinfo {author} {\bibfnamefont {A.}~\bibnamefont
  {Hearmon}}, \bibinfo {author} {\bibfnamefont {F.}~\bibnamefont {Fabrizi}},
  \bibinfo {author} {\bibfnamefont {P.~G.}\ \bibnamefont {Radaelli}}, \bibinfo
  {author} {\bibfnamefont {S.-W.}\ \bibnamefont {Cheong}}, \ and\ \bibinfo
  {author} {\bibfnamefont {L.~C.}\ \bibnamefont {Chapon}},\ }\href {\doibase
  10.1103/PhysRevB.102.054417} {\bibfield  {journal} {\bibinfo  {journal}
  {Phys. Rev. B}\ }\textbf {\bibinfo {volume} {102}},\ \bibinfo {pages}
  {054417} (\bibinfo {year} {2020})}\BibitemShut {NoStop}%
\bibitem [{\citenamefont {{Giles-Donovan}}\ \emph {et~al.}(2020)\citenamefont
  {{Giles-Donovan}}, \citenamefont {Qureshi}, \citenamefont {Johnson},
  \citenamefont {Zhang}, \citenamefont {Cheong}, \citenamefont {Cochran},\ and\
  \citenamefont {Stock}}]{giles-donovan2020102}%
  \BibitemOpen
  \bibfield  {author} {\bibinfo {author} {\bibfnamefont {N.}~\bibnamefont
  {{Giles-Donovan}}}, \bibinfo {author} {\bibfnamefont {N.}~\bibnamefont
  {Qureshi}}, \bibinfo {author} {\bibfnamefont {R.~D.}\ \bibnamefont
  {Johnson}}, \bibinfo {author} {\bibfnamefont {L.~Y.}\ \bibnamefont {Zhang}},
  \bibinfo {author} {\bibfnamefont {S.-W.}\ \bibnamefont {Cheong}}, \bibinfo
  {author} {\bibfnamefont {S.}~\bibnamefont {Cochran}}, \ and\ \bibinfo
  {author} {\bibfnamefont {C.}~\bibnamefont {Stock}},\ }\href {\doibase
  10.1103/PhysRevB.102.024414} {\bibfield  {journal} {\bibinfo  {journal}
  {Phys. Rev. B}\ }\textbf {\bibinfo {volume} {102}},\ \bibinfo {pages}
  {024414} (\bibinfo {year} {2020})}\BibitemShut {NoStop}%
\bibitem [{\citenamefont {Brown}(2006)}]{BROWN2006215}%
  \BibitemOpen
  \bibfield  {author} {\bibinfo {author} {\bibfnamefont {P.~J.}~\bibnamefont
  {Brown}},\ }in\ \href {\doibase 10.1016/B978-044451050-1/50006-9} {\emph
  {\bibinfo {booktitle} {Neutron Scattering from Magnetic Materials}}},\
  \bibinfo {series and number} {\bibinfo {number} {Chap. 5}},\ \bibinfo
  {editor} {edited by\ \bibinfo {editor} {\bibfnamefont {T.}~\bibnamefont
  {Chatterji}}}\ (\bibinfo  {publisher} {{Elsevier Science}},\ \bibinfo
  {address} {{Amsterdam}},\ \bibinfo {year} {2006})\ pp.\ \bibinfo {pages}
  {215--244}\BibitemShut {NoStop}%
\bibitem [{\citenamefont {Simonet}\ \emph {et~al.}(2012)\citenamefont
  {Simonet}, \citenamefont {Loire},\ and\ \citenamefont
  {Ballou}}]{simonet2012213}%
  \BibitemOpen
  \bibfield  {author} {\bibinfo {author} {\bibfnamefont {V.}~\bibnamefont
  {Simonet}}, \bibinfo {author} {\bibfnamefont {M.}~\bibnamefont {Loire}}, \
  and\ \bibinfo {author} {\bibfnamefont {R.}~\bibnamefont {Ballou}},\ }\href
  {\doibase 10.1140/epjst/e2012-01661-8} {\bibfield  {journal} {\bibinfo
  {journal} {Eur. Phys. J.: Spec. Top.}\ }\textbf {\bibinfo {volume} {213}},\
  \bibinfo {pages} {5} (\bibinfo {year} {2012})}\BibitemShut {NoStop}%
\bibitem [{\citenamefont {Blume}(1963)}]{blume1963130}%
  \BibitemOpen
  \bibfield  {author} {\bibinfo {author} {\bibfnamefont {M.}~\bibnamefont
  {Blume}},\ }\href {\doibase 10.1103/PhysRev.130.1670} {\bibfield  {journal}
  {\bibinfo  {journal} {Phys. Rev.}\ }\textbf {\bibinfo {volume} {130}},\
  \bibinfo {pages} {1670} (\bibinfo {year} {1963})}\BibitemShut {NoStop}%
\bibitem [{\citenamefont {Maleev}\ \emph {et~al.}(1963)\citenamefont {Maleev},
  \citenamefont {Bar'yakhtar},\ and\ \citenamefont {Suris}}]{maleev19634}%
  \BibitemOpen
  \bibfield  {author} {\bibinfo {author} {\bibfnamefont {S.~V.}\ \bibnamefont
  {Maleev}}, \bibinfo {author} {\bibfnamefont {V.~G.}\ \bibnamefont
  {Bar'yakhtar}}, \ and\ \bibinfo {author} {\bibfnamefont {R.~A.}\ \bibnamefont
  {Suris}},\ }\href {https://www.osti.gov/biblio/4713671} {\bibfield  {journal}
  {\bibinfo  {journal} {Sov. Phys. Solid State}\ }\textbf {\bibinfo {volume} {4}}, \bibinfo {pages}{2533}
  (\bibinfo {year} {1963})}\BibitemShut {NoStop}%
\bibitem [{\citenamefont {Hahn}\ and\ \citenamefont
  {Klapper}(2013)}]{hahn2013D}%
  \BibitemOpen
  \bibfield  {author} {\bibinfo {author} {\bibfnamefont {T.}~\bibnamefont
  {Hahn}}\ and\ \bibinfo {author} {\bibfnamefont {H.}~\bibnamefont {Klapper}},\
  }in\ \href {\doibase 10.1107/97809553602060000917} {\emph {\bibinfo
  {booktitle} {International {{Tables}} for {{Crystallography}}}}},\
\bibinfo {editor} {edited by\ \bibinfo {editor}
  {\bibfnamefont {A.}~\bibnamefont {Authier,}}}\, 2nd ed.
  (\bibinfo {year} {Wiley, 2013}),\  Vol.~\bibinfo {volume} {D},
 pp.\ \bibinfo {pages} {413--487}\BibitemShut
  {NoStop}%
\bibitem [{\citenamefont {Parsons}(2003)}]{parsons200359}%
  \BibitemOpen
  \bibfield  {author} {\bibinfo {author} {\bibfnamefont {S.}~\bibnamefont
  {Parsons}},\ }\href {\doibase 10.1107/S0907444903017657} {\bibfield
  {journal} {\bibinfo  {journal} {Acta Crystallogr. D}\ }\textbf {\bibinfo
  {volume} {59}},\ \bibinfo {pages} {1995} (\bibinfo {year}
  {2003})}\BibitemShut {NoStop}%
\bibitem [{\citenamefont {Koch}(2006)}]{koch2006C}%
  \BibitemOpen
  \bibfield  {author} {\bibinfo {author} {\bibfnamefont {E.}~\bibnamefont
  {Koch}},\ }in\ \href {\doibase 10.1107/97809553602060000572} {\emph {\bibinfo
  {booktitle} {International {{Tables}} for {{Crystallography}}}}},\
   \bibinfo {editor} {edited by\ \bibinfo {editor}
  {\bibfnamefont {E.}~\bibnamefont {Prince,}}}\ 
  \bibinfo {edition} {1st}\ ed.\
  (\bibinfo {year} {Wiley, 2006}),\
   Vol.~\bibinfo {volume} {C},\ pp.\ \bibinfo {pages} {10--14}\BibitemShut
  {NoStop}%
\bibitem [{\citenamefont {Chandra}\ \emph {et~al.}(1999)\citenamefont
  {Chandra}, \citenamefont {Acharya},\ and\ \citenamefont
  {Moody}}]{chandra199955}%
  \BibitemOpen
  \bibfield  {author} {\bibinfo {author} {\bibfnamefont {N.}~\bibnamefont
  {Chandra}}, \bibinfo {author} {\bibfnamefont {K.~R.}\ \bibnamefont
  {Acharya}}, \ and\ \bibinfo {author} {\bibfnamefont {P.~C.~E.}\ \bibnamefont
  {Moody}},\ }\href {\doibase 10.1107/S0907444999009968} {\bibfield  {journal}
  {\bibinfo  {journal} {Acta Crystallogr. D}\ }\textbf {\bibinfo {volume}
  {55}},\ \bibinfo {pages} {1750} (\bibinfo {year} {1999})}\BibitemShut
  {NoStop}%
\bibitem [{\citenamefont {Flack}(2003)}]{flack200386}%
  \BibitemOpen
  \bibfield  {author} {\bibinfo {author} {\bibfnamefont {H.~D.}\ \bibnamefont
  {Flack}},\ }\href {\doibase 10.1002/hlca.200390109} {\bibfield  {journal}
  {\bibinfo  {journal} {Helv. Chim. Acta}\ }\textbf {\bibinfo {volume} {86}},\
  \bibinfo {pages} {905} (\bibinfo {year} {2003})}\BibitemShut {NoStop}%
\bibitem [{\citenamefont {Barron}(1986)}]{barron1986108}%
  \BibitemOpen
  \bibfield  {author} {\bibinfo {author} {\bibfnamefont {L.~D.}\ \bibnamefont
  {Barron}},\ }\href {\doibase 10.1021/ja00278a029} {\bibfield  {journal}
  {\bibinfo  {journal} {J. Am. Chem. Soc.}\ }\textbf {\bibinfo {volume}
  {108}},\ \bibinfo {pages} {5539} (\bibinfo {year} {1986})}\BibitemShut
  {NoStop}%
\bibitem [{\citenamefont {Johnson}\ and\ \citenamefont
  {Radaelli}(2014)}]{johnson201444}%
  \BibitemOpen
  \bibfield  {author} {\bibinfo {author} {\bibfnamefont {R.~D.}\ \bibnamefont
  {Johnson}}\ and\ \bibinfo {author} {\bibfnamefont {P.~G.}\ \bibnamefont
  {Radaelli}},\ }\href {\doibase 10.1146/annurev-matsci-070813-113524}
  {\bibfield  {journal} {\bibinfo  {journal} {Annu. Rev. Mater. Res.}\ }\textbf
  {\bibinfo {volume} {44}},\ \bibinfo {pages} {269} (\bibinfo {year}
  {2014})}\BibitemShut {NoStop}%
\bibitem [{\citenamefont {Villain}(1977)}]{villain197710}%
  \BibitemOpen
  \bibfield  {author} {\bibinfo {author} {\bibfnamefont {J.}~\bibnamefont
  {Villain}},\ }\href {\doibase 10.1088/0022-3719/10/23/013} {\bibfield
  {journal} {\bibinfo  {journal} {J. Phys. Condens. Matter}\ }\textbf {\bibinfo
  {volume} {10}},\ \bibinfo {pages} {4793} (\bibinfo {year}
  {1977})}\BibitemShut {NoStop}%
\bibitem [{\citenamefont
  {{Rodr{\'i}guez-Carvajal}}(1993)}]{rodriguez-carvajal1993192}%
  \BibitemOpen
  \bibfield  {author} {\bibinfo {author} {\bibfnamefont {J.}~\bibnamefont
  {{Rodr{\'i}guez-Carvajal}}},\ }\href {\doibase 10.1016/0921-4526(93)90108-I}
  {\bibfield  {journal} {\bibinfo  {journal} {Phys. B}\ }\textbf {\bibinfo
  {volume} {192}},\ \bibinfo {pages} {55} (\bibinfo {year} {1993})}\BibitemShut
  {NoStop}%
\bibitem [{\citenamefont {Jerphagnon}\ and\ \citenamefont
  {Chemla}(1976)}]{jerphagnon197665}%
  \BibitemOpen
  \bibfield  {author} {\bibinfo {author} {\bibfnamefont {J.}~\bibnamefont
  {Jerphagnon}}\ and\ \bibinfo {author} {\bibfnamefont {D.~S.}\ \bibnamefont
  {Chemla}},\ }\href {\doibase 10.1063/1.433207} {\bibfield  {journal}
  {\bibinfo  {journal} {J. Chem. Phys.}\ }\textbf {\bibinfo {volume} {65}},\
  \bibinfo {pages} {1522} (\bibinfo {year} {1976})}\BibitemShut {NoStop}%
\bibitem [{\citenamefont {Wang}\ \emph {et~al.}(2015)\citenamefont {Wang},
  \citenamefont {Huang}, \citenamefont {Yang}, \citenamefont {Oh},\ and\
  \citenamefont {Cheong}}]{wang20153}%
  \BibitemOpen
  \bibfield  {author} {\bibinfo {author} {\bibfnamefont {X.}~\bibnamefont
  {Wang}}, \bibinfo {author} {\bibfnamefont {F.-T.}\ \bibnamefont {Huang}},
  \bibinfo {author} {\bibfnamefont {J.}~\bibnamefont {Yang}}, \bibinfo {author}
  {\bibfnamefont {Y.~S.}\ \bibnamefont {Oh}}, \ and\ \bibinfo {author}
  {\bibfnamefont {S.-W.}\ \bibnamefont {Cheong}},\ }\href {\doibase
  10.1063/1.4927232} {\bibfield  {journal} {\bibinfo  {journal} {APL Mater.}\
  }\textbf {\bibinfo {volume} {3}},\ \bibinfo {pages} {076105} (\bibinfo {year}
  {2015})}\BibitemShut {NoStop}%
\bibitem [{\citenamefont {Prosnikov}\ \emph {et~al.}(2019)\citenamefont
  {Prosnikov}, \citenamefont {Smirnov}, \citenamefont {Davydov}, \citenamefont
  {Araki}, \citenamefont {Arima},\ and\ \citenamefont
  {Pisarev}}]{prosnikov2019100}%
  \BibitemOpen
  \bibfield  {author} {\bibinfo {author} {\bibfnamefont {M.~A.}\ \bibnamefont
  {Prosnikov}}, \bibinfo {author} {\bibfnamefont {A.~N.}\ \bibnamefont
  {Smirnov}}, \bibinfo {author} {\bibfnamefont {V.~Y.}\ \bibnamefont
  {Davydov}}, \bibinfo {author} {\bibfnamefont {Y.}~\bibnamefont {Araki}},
  \bibinfo {author} {\bibfnamefont {T.}~\bibnamefont {Arima}}, \ and\ \bibinfo
  {author} {\bibfnamefont {R.~V.}\ \bibnamefont {Pisarev}},\ }\href {\doibase
  10.1103/PhysRevB.100.144417} {\bibfield  {journal} {\bibinfo  {journal}
  {Phys. Rev. B}\ }\textbf {\bibinfo {volume} {100}},\ \bibinfo {pages}
  {144417} (\bibinfo {year} {2019})}\BibitemShut {NoStop}%
\bibitem [{\citenamefont {Campostrini}\ \emph {et~al.}(2002)\citenamefont
  {Campostrini}, \citenamefont {Hasenbusch}, \citenamefont {Pelissetto},
  \citenamefont {Rossi},\ and\ \citenamefont {Vicari}}]{campostrini200265}%
  \BibitemOpen
  \bibfield  {author} {\bibinfo {author} {\bibfnamefont {M.}~\bibnamefont
  {Campostrini}}, \bibinfo {author} {\bibfnamefont {M.}~\bibnamefont
  {Hasenbusch}}, \bibinfo {author} {\bibfnamefont {A.}~\bibnamefont
  {Pelissetto}}, \bibinfo {author} {\bibfnamefont {P.}~\bibnamefont {Rossi}}, \
  and\ \bibinfo {author} {\bibfnamefont {E.}~\bibnamefont {Vicari}},\ }\href
  {\doibase 10.1103/PhysRevB.65.144520} {\bibfield  {journal} {\bibinfo
  {journal} {Phys. Rev. B}\ }\textbf {\bibinfo {volume} {65}},\ \bibinfo
  {pages} {144520} (\bibinfo {year} {2002})}\BibitemShut {NoStop}%
\bibitem [{\citenamefont {Kawamura}(1988)}]{kawamura198863}%
  \BibitemOpen
  \bibfield  {author} {\bibinfo {author} {\bibfnamefont {H.}~\bibnamefont
  {Kawamura}},\ }\href {\doibase 10.1063/1.340905} {\bibfield  {journal}
  {\bibinfo  {journal} {J. App. Phys.}\ }\textbf {\bibinfo {volume} {63}},\
  \bibinfo {pages} {3086} (\bibinfo {year} {1988})}\BibitemShut {NoStop}%
\bibitem [{\citenamefont {Qureshi}\ \emph {et~al.}(2009)\citenamefont
  {Qureshi}, \citenamefont {Zbiri}, \citenamefont {Rodr\'{\i}guez-Carvajal},
  \citenamefont {Stunault}, \citenamefont {Ressouche}, \citenamefont {Hansen},
  \citenamefont {Fern\'andez-D\'{\i}az}, \citenamefont {Johnson}, \citenamefont
  {Fuess}, \citenamefont {Ehrenberg}, \citenamefont {Sakurai}, \citenamefont
  {Itou}, \citenamefont {Gillon}, \citenamefont {Wolf}, \citenamefont
  {Rodr\'{\i}guez-Velamazan},\ and\ \citenamefont
  {S\'anchez-Montero}}]{qureshi200979}%
  \BibitemOpen
  \bibfield  {author} {\bibinfo {author} {\bibfnamefont {N.}~\bibnamefont
  {Qureshi}}, \bibinfo {author} {\bibfnamefont {M.}~\bibnamefont {Zbiri}},
  \bibinfo {author} {\bibfnamefont {J.}~\bibnamefont
  {Rodr\'{\i}guez-Carvajal}}, \bibinfo {author} {\bibfnamefont
  {A.}~\bibnamefont {Stunault}}, \bibinfo {author} {\bibfnamefont
  {E.}~\bibnamefont {Ressouche}}, \bibinfo {author} {\bibfnamefont {T.~C.}\
  \bibnamefont {Hansen}}, \bibinfo {author} {\bibfnamefont {M.~T.}\
  \bibnamefont {Fern\'andez-D\'{\i}az}}, \bibinfo {author} {\bibfnamefont
  {M.~R.}\ \bibnamefont {Johnson}}, \bibinfo {author} {\bibfnamefont
  {H.}~\bibnamefont {Fuess}}, \bibinfo {author} {\bibfnamefont
  {H.}~\bibnamefont {Ehrenberg}}, \bibinfo {author} {\bibfnamefont
  {Y.}~\bibnamefont {Sakurai}}, \bibinfo {author} {\bibfnamefont
  {M.}~\bibnamefont {Itou}}, \bibinfo {author} {\bibfnamefont {B.}~\bibnamefont
  {Gillon}}, \bibinfo {author} {\bibfnamefont {T.}~\bibnamefont {Wolf}},
  \bibinfo {author} {\bibfnamefont {J.~A.}\ \bibnamefont
  {Rodr\'{\i}guez-Velamazan}}, \ and\ \bibinfo {author} {\bibfnamefont
  {J.}~\bibnamefont {S\'anchez-Montero}},\ }\href {\doibase
  10.1103/PhysRevB.79.094417} {\bibfield  {journal} {\bibinfo  {journal} {Phys.
  Rev. B}\ }\textbf {\bibinfo {volume} {79}},\ \bibinfo {pages} {094417}
  (\bibinfo {year} {2009})}\BibitemShut {NoStop}%
\bibitem [{\citenamefont {Yosida}(1996)}]{Yosida:book}%
  \BibitemOpen
  \bibfield  {author} {\bibinfo {author} {\bibfnamefont {K.}~\bibnamefont
  {Yosida}},\ }\href@noop {} {\emph {\bibinfo {title} {Theory of Magnetism}}}\
  (\bibinfo  {publisher} {Springer},\ \bibinfo {address} {New York},\ \bibinfo
  {year} {1996})\BibitemShut {NoStop}%
\end{thebibliography}
\end{document}